\documentclass[twocolumn, times]{aastex7}
\usepackage{amsmath}
\usepackage{booktabs}

\newcommand{\cxo}{\textrm{Chandra}}
\newcommand{\herschel}{\textrm{Herschel}}
\newcommand{\astrosat}{\textrm{AstroSat}}
\newcommand{\spitzer}{\textrm{Spitzer}}

\newcommand{\psfsize}{$25''$}
\newcommand{\pixsize}{$10''$}

\newcommand{\physsize}{$0.403~{\rm kpc^2~pix^{-1}}$}

\newcommand{\nbins}{7}

\newcommand{\nsteps}{20000}
\newcommand{\nwalkers}{64}
\newcommand{\ndiscard}{5000}
\newcommand{\nthin}{500}

\newcommand{\galctrra}{184.707}
\newcommand{\galctrdec}{14.417}

\newcommand{\angextent}{$65\farcs6^{+1\farcs9}_{-2\farcs1}$}
\newcommand{\physextent}{$4.2\pm0.1$}
\newcommand{\kTextent}{$0.19_{-0.04}^{+0.06}$}
\newcommand{\LXSBabsextent}{$3.63^{+0.18}_{-0.21} \times 10^{39}$}
\newcommand{\LXSBunabsextent}{$1.52^{+2.34}_{-0.87} \times 10^{40}$}

\newcommand{\kTextenttwo}{$0.15_{-0.03}^{+0.06}$}

\newcommand{\SFRD}{\ensuremath{\Sigma_{\rm SFR}}}

\newcommand{\acisx}{\texttt{ACIS~Extract}}
\newcommand{\lgh}{\texttt{Lightning}}
\newcommand{\cloudy}{\texttt{Cloudy}}

\defcitealias{chevalier1985}{CC85}
\defcitealias{draine2007}{DL07}

\begin{document}

\title{Constraining the Sub-Galactic Relationship Between Star Formation and the Hot Interstellar Medium in NGC 4254}

\correspondingauthor{Erik B. Monson}

\author[0000-0001-8473-5140]{Erik B. Monson}
\affiliation{Department of Physics and Astronomy, Middle Tennessee State University, 1301 E. Main Street Box 71, Murfreesboro, TN 37132, USA}
\affiliation{Department of Astronomy and Astrophysics, Pennsylvania State University, 525 Davey Lab, University Park, PA 16802, USA}
\email{erik.monson@mtsu.edu}

\author[0000-0003-2192-3296]{Bret D. Lehmer}
\affiliation{Department of Physics, University of Arkansas, 226 Physics Building, 825 West Dickson Street, Fayetteville, AR 72701, USA}
\email{lehmer@uark.edu}

\author[0000-0002-8553-1964]{Amirnezam Amiri}
\affiliation{Department of Physics, University of Arkansas, 226 Physics Building, 825 West Dickson Street, Fayetteville, AR 72701, USA}
\email{aa219@uark.edu}


\author[0000-0003-4660-9762]{Karina Barboza}
\affiliation{Department of Astronomy, The Ohio State University, 140 W. 18th Ave., Columbus, OH 43210, USA}
\affiliation{Center for Cosmology and AstroParticle Physics, The Ohio State University, 191 W. Woodruff Ave., Columbus, OH 43210, USA}
\email{barboza.21@osu.edu}

\author[0000-0003-0410-4504]{Ashley~T.~Barnes}
\affiliation{European Southern Observatory, Karl-Schwarzschild-Strasse 2, D-85748 Garching bei München, Germany}
\email{ashleybarnes.astro@gmail.com}

\author[0000-0001-8525-4920]{Antara~R.~Basu-Zych}
\affiliation{Department of Physics, University of Maryland Baltimore County, Baltimore, MD 21250, USA}
\affiliation{NASA Goddard Space Flight Center, Code 662, Greenbelt, MD 20771, USA}
\affiliation{Center for Research and Exploration in Space Science and Technology, NASA/GSFC, Greenbelt, MD 20771, USA}
\email{antara.r.basu-zych@nasa.gov}

\author[0000-0002-5782-9093]{Daniel~A.~Dale}
\affiliation{Department of Physics and Astronomy, University of Wyoming, Laramie, WY 82071, USA}
\email{ddale@uwyo.edu}

\author[0000-0002-9069-7061]{Sanskriti Das}
\altaffiliation{NASA Hubble Fellow}
\affiliation{Kavli Institute for Particle Astrophysics and Cosmology, Stanford University, 452 Lomita Mall, Stanford, CA\,94305, USA}
\email{snskriti@stanford.edu}

\author[0000-0002-2885-6172]{Simthembile Dlamini}
\affiliation{Department of Astronomy, University of Cape Town, Rondebosch 7701, South Africa}
\email{simther4111@gmail.com}

\author[0000-0001-6708-1317]{Simon Glover}
\affiliation{Institute for Theoretical Astrophysics, University of Heidelberg, Albert-Ueberle-Strasse 2, 69120 Heidelberg, Germany}
\email{glover@uni-heidelberg.de}

\author[0000-0001-6551-3091]{Kathryn Kreckel}
\affiliation{Astronomisches Rechen-Institut, Zentrum f\"{u}r Astronomie der Universit\"{a}t Heidelberg, M\"{o}nchhofstra\ss e 12-14, D-69120 Heidelberg, Germany}
\email{kathryn.kreckel@uni-heidelberg.de}

\author[0000-0002-1790-3148]{Laura A. Lopez}
\affiliation{Department of Astronomy, The Ohio State University, 140 W. 18th Ave., Columbus, OH 43210, USA}
\affiliation{Center for Cosmology and AstroParticle Physics, The Ohio State University, 191 W. Woodruff Ave., Columbus, OH 43210, USA}
\email{lopez.513@osu.edu}

\author[0000-0002-2644-0077]{Sebastian Lopez} 
\affiliation{Department of Astronomy, The Ohio State University, 140 W. 18th Ave., Columbus, OH 43210, USA}
\affiliation{Center for Cosmology and AstroParticle Physics, The Ohio State University, 191 W. Woodruff Ave., Columbus, OH 43210, USA}
\email{lopez.764@osu.edu}

\author[0000-0002-4822-3559]{Smita Mathur}
\affiliation{Department of Astronomy, The Ohio State University, 140 W. 18th Ave., Columbus, OH 43210, USA}
\affiliation{Center for Cosmology and AstroParticle Physics, The Ohio State University, 191 W. Woodruff Ave., Columbus, OH 43210, USA}
\email{mathur.17@osu.edu}

\author[0000-0002-1370-6964]{Hsi-An Pan}
\affiliation{Department of Physics, Tamkang University, No.151, Yingzhuan Road, Tamsui District, New Taipei City 251301, Taiwan} 
\email{hapan@gms.tku.edu.tw}

\author[0000-0003-1560-001X]{Jennifer A. Rodriguez}
\affiliation{Department of Astronomy, The Ohio State University, 140 W. 18th Ave., Columbus, OH 43210, USA}
\affiliation{Center for Cosmology and AstroParticle Physics, The Ohio State University, 191 W. Woodruff Ave., Columbus, OH 43210, USA}
\email{rodriguez.1392@osu.edu}

\author[0000-0002-4378-8534]{Karin Sandstrom}
\affiliation{Department of Astronomy \& Astrophysics, University of California, San Diego, 9500 Gilman Drive MC0424, La Jolla, CA 92093, USA}
\email{kmsandstrom@ucsd.edu}

\author[0000-0002-6313-4597]{Sumit K. Sarbadhicary}
\affiliation{Department of Physics and Astronomy, The Johns Hopkins University, Baltimore, MD 21218 USA}
\email{ssarbad1@jh.edu}

\author[0000-0003-0378-4667]{Jiayi~Sun}
\altaffiliation{NASA Hubble Fellow}
\affiliation{Department of Astrophysical Sciences, Princeton University, 4 Ivy Lane, Princeton, NJ 08544, USA}
\affiliation{Department of Physics and Astronomy, University of Kentucky, 506 Library Drive, Lexington, KY 40506, USA}
\email{jiayi.sun@uky.edu}

\author[0000-0002-0012-2142]{Thomas~G.~Williams}
\affiliation{Sub-department of Astrophysics, Department of Physics, University of Oxford, Keble Road, Oxford OX1 3RH, UK}
\email{thomas.williams@physics.ox.ac.uk}

\begin{abstract}

We investigate the relationship between star formation and X-ray emission from the hot interstellar medium (ISM) on $\sim$kpc scales in NGC 4254 (M99) by combining spatially resolved star formation histories (SFHs) and Bayesian X-ray spectral fitting.
We measure sub-galactic star formation rates (SFR) by modeling spectrophotometric UV-IR data with flexible SFHs, and we produce point-source-subtracted maps of the diffuse X-ray emission using Chandra data.
We extract and fit the spectra of 5 regions selected by their SFR density \SFRD, deriving hot gas luminosities and plasma temperatures. We examine the sub-galactic $kT-\SFRD$ and $L^{\rm gas}_X-\SFRD$ scaling relations in NGC 4254, and compare to predictions from simple models of the feedback into the ISM from core collapse supernovae (CCSNe).
The hot gas emission from NGC 4254 is consistent with thermalization of $\approx 40-50\%$ of the energy from CCSNe in the ISM, and mass-loading of the CCSNe ejecta which decreases as $\SFRD^{-1/3}$.
Our optimized model implies a temperature and X-ray production efficiency that scale as $kT = (0.72^{+0.26}_{-0.18}~{\rm keV}) \SFRD^{0.34\pm0.10}$ and $\eta = (0.03^{+0.02}_{-0.01}) \SFRD^{0.34\pm0.18}$, respectively, for $\SFRD = 0.01-0.13~{\rm M_{\odot}~yr^{-1}~kpc^{-2}}$.
We also compare the properties of the hot ISM to other ISM phases using data from the PHANGS program. The diffuse X-ray emission of a given region is on average 200 times fainter than the H$\alpha$ emission, and we see evidence that the hot ISM is over-pressurized compared to the large-scale dynamical equilibrium pressure of the galaxy, consistent with expansion of the hot ISM into the ambient medium.

\end{abstract}


\section{Introduction} \label{sec:intro}

Since the Einstein mission, normal galaxies (i.e., with negligible nuclear activity from accreting supermassive black holes) have been recognized as extended X-ray sources, with significant diffuse, soft ($\lesssim 2$ keV) emission due to a hot ($\sim10^6-10^7$ K) phase of the interstellar medium (ISM), heated by shocks from supernova-driven winds \citep[see e.g. review by][]{fabbiano1989}. The advent of high-resolution X-ray astronomy with Chandra allowed extensive separation of resolved X-ray point source populations from the diffuse ISM emission beyond galaxies in the Local Group (e.g. \citealp{mineo2012}; see also \citealp{fabbiano2019} for a review of Chandra's impact on the field), such that the isolated emission from the ISM could be correlated with galaxy properties, conclusively demonstrating the link between star formation and the X-ray-emitting ISM \citep[see][for a recent review]{nardini2022}. The galaxy-integrated star formation rate (SFR) and the integrated X-ray luminosity from the hot ISM are observed to follow a relatively tight linear correlation over three orders of magnitude in luminosity and SFR: \added{\citet{mineo2012} found $L_X /{\rm SFR} = 5.2\times10^{38}~{\rm erg~s^{-1}~(M_{\odot}~yr^{-1})^{-1}}$ with only 0.34 dex scatter}.

In addition to kinetic energy, ejecta from core-collapse supernovae (CCSNe) contain metals synthesized in the progenitor star, enriching the hot ISM \citep[e.g.][]{baldi2006, nardini2013}. If the velocity of the wind exceeds the escape velocity of the galaxy, energy and baryons can be removed from the galaxy in an outflow \citep[e.g.][]{heckman1990}, potentially enriching the circumgalactic and intergalactic media with metals \citep{mathur2022}. Observations of the diffuse hot gas emission in X-rays can thus directly probe the baryon cycle in galaxies and constrain supernova-driven wind models \citep[e.g][]{zhang2014, meiksin2016, thompson2016}. Models of feedback from supernova winds are an important component of cosmological simulations, which require feedback from supernova and stellar winds to prevent runaway star formation and produce realistic galaxies \citep[e.g.][]{hopkins2012, hopkins2014}. Since stellar and supernova feedback operate on spatial scales below the resolution of most large-scale simulations, they are implemented as ``subgrid'' processes and relying on empirical constraints \citetext{see review by \citealp{naab2017} and updates by \citealp{schneider2020}}. A better understanding of the parameter space of supernova-driven winds will thus also improve our ability to simulate realistic galaxies and galaxy populations.


One of the simplest models for the structure of a high-temperature plasma shock-heated by winds from supernovae was developed by \citet{chevalier1985} (CC85, hereafter).
\citetalias{chevalier1985} models an adiabatically expanding, spherically symmetric wind, 
launched from a spherically symmetric star-forming region with a size on the order of $R \approx 100$ pc.
Inside this region, mass and energy are deposited as a steady-state process by CCSNe. The supernova ejecta shock-heats the ISM and entrains additional mass into the ISM (see \autoref{sec:cc85} for details). In this model, the hot ISM consists of a single phase (i.e., a single temperature), neglecting the interaction between the hot and cold phases of the ISM \citep[see][for a recent example of a multiphase model]{fielding2022}. The radiative cooling of the wind is neglected, assuming that the wind cools primarily by adiabatic expansion.
The \citetalias{chevalier1985} model has been applied to study the radial structure of superwinds in several canonical outflowing, edge-on starburst galaxies: e.g., M82 \citep{strickland2009, lopez2020}, NGC 253 \citep{lopez2023}, and NGC 4945 \citep{porrazbarrera2024}. These studies have observed that the central structure of the outflow, nearest the wind-launching region, can be approximated with this simple model. However, the temperature and density profiles of the wind depart from the model expectation at large altitudes above the disk. \added{In this work we instead use the \citetalias{chevalier1985} model to interpret the diffuse X-ray emission in the plane of a face-on galaxy. We treat sub-galactic regions as sites of star formation launching adiabatically-expanding winds into the galaxy, and investigate variations in the X-ray production efficiency and mass-loading of the X-ray plasma as a function of star formation rate.}

Constraining the relationship between the soft X-ray emission and SFR on sub-galactic scales \citep[e.g][]{yukita2010, yukita2012, kouroumpatzakis2020, zhang2024, zhang2025} imposes two main requirements. \added{We first require deep ($\gtrsim 50-100$ ks) X-ray observations, which can permit the identification and masking of point sources, and collect large numbers of photons associated with diffuse emission per kpc-scale region to constrain the temperature and luminosity of the X-ray plasma. We also need observations of a given galaxy in UV, optical, and IR bands to robustly constrain the sub-galactic SFR, either with calibrated SFR scaling relations or by forward-modeling the photometry with population synthesis models and a flexible star-formation history (SFH) model.} These sub-galactic studies occasionally reveal departures from the global scaling, as in \citet{zhang2024} and \citet{zhang2025}, where the nuclear regions of several star-forming galaxies are seen to exhibit steeper-than-linear scaling relations, or \citet{kouroumpatzakis2020}, where low-SFR regions are seen to exhibit an X-ray excess attributed to un-subtracted low-mass X-ray binaries. 

NGC 4254 (M99; R.A. = \galctrra\ deg, Dec. = \galctrdec\ deg, \added{z=0.008; \citealp{lang2020}}) is a nearby (13.1 Mpc; \citealp{anand2021, nugent2006}), low-inclination ($i=34.4$ deg; \citealp{lang2020}) spiral galaxy with a three-arm structure, $\log M_{\star} /{\rm M_{\odot}} = 10.42$, and $\log {\rm SFR / (M_{\odot}~yr^{-1})} = 0.49$ \citep{leroy2021a}. \added{At this distance the physical scale is 0.064 kpc arcsec$^{-1}$.} The galaxy is thought to have been tidally disturbed by interactions within the Virgo Cluster, producing enhanced star formation compared to other nearby spirals \citep{chyzy2007}. As part of the Physics at High Angular Resolution in Nearby GalaxieS (PHANGS) sample, \added{NGC 4254 meets or exceeds our stated requirements for sub-galactic constraints on the relationship between the soft X-ray emission and SFR.}
NGC 4254 has $\sim 100$ ks of data in the \cxo\ archive, with $\sim 50$ ks obtained in 2007 and 2015, when \cxo\ retained significant sensitivity to soft X-rays below 1 keV. \added{These data allow us to map the properties of the hot phase of the ISM}. 
\added{The galaxy also has a wealth of archival imaging spanning the UV to far-IR, allowing robust sub-galactic constraints on the SFR.} 
\added{In addition, new observations from} the PHANGS survey provide unique, high spatial resolution access to the molecular and warm ionized phases of the ISM with CO(2-1) mapping from the Atacama Large Millimeter Array (ALMA) and integral field unit (IFU) spectroscopic coverage from the Multi-Unit Spectroscopic Explorer (MUSE) on the Very Large Telescope (VLT).
In what follows, we use NGC 4254 to study sub-galactic variations in the X-ray emitting ISM and interpret them in the context of the \citetalias{chevalier1985} model \added{and the multi-phase ISM traced by PHANGS}, as a proof-of-concept \added{and a pathfinder} for how much constraint a single \added{low-inclination} PHANGS galaxy with moderate \added{X-ray} observing depth can place on the parameters of simple supernova wind models and the relationship between the different phases of the ISM. 
\added{Future work will expand the analyses presented below to the remainder of the PHANGS-Chandra survey, providing excellent constraints on supernova wind models and kpc-scale correlations of the multi-phase ISM in a diverse sample of galactic environments.}

The paper is organized as follows. In \autoref{sec:data} we describe the datasets we utilize; in \autoref{sec:sfh} we describe our methods for deriving robust spatially-resolved SFHs and SN rates from SED fitting; in \autoref{sec:spec} we describe our methods for deriving spatially-resolved X-ray luminosities and plasma temperatures using Bayesian X-ray spectral fitting; in \autoref{sec:discussion} we examine the sub-galactic relationship between X-ray emission and star formation, interpret our results in the context of the \citetalias{chevalier1985} model solution described in \autoref{sec:cc85}, and examine the relationship between different phases of the ISM. Finally, \autoref{sec:summary} provides a summary of our findings.

\section{Data and Reduction} \label{sec:data}

\subsection{\cxo\ X-ray Data} \label{sec:data:xray}

The \cxo\ observations listed in \autoref{table:obslog} were retrieved from the \cxo\ science archive\footnote{\url{https://cda.harvard.edu/chaser/}} and reduced using standard \texttt{CIAO} v4.16 tools and \texttt{CALDB} v4.11.2. We used the \verb|chandra_repro| script to generate Level 2 products, then constructed a background light curve and filtered out flares with \texttt{deflare}. We projected the filtered event lists to the common frame of the deepest observation (ObsID 17462) using \verb|wcs_match| and \verb|wcs_update|. Typical shifts are on the order of $1-3$ sky pixels ($\approx 0.5-1.5''$); the X and Y image-plane shifts are given for each observation in \autoref{table:obslog}\footnote{\added{Careful reprojection of the Chandra data to the astrometric frame of the PHANGS data described in \autoref{sec:data:phangs} performed by matching Chandra- and JWST-detected point sources will be presented in Lehmer et al. (2026, in prep.). On a preliminary basis, we estimate that the offset between the Chandra frame described here and the PHANGS frame is $\approx0\farcs24\pm0\farcs10$ ($\approx 0.5$ Chandra pixels), significantly smaller than the kpc scales on which we model the X-ray data and ISM correlations in what follows. We thus do not apply further absolute astrometric corrections beyond correcting the relative astrometry of the Chandra observations.}} We generated exposure maps with \texttt{fluximage} and produced a preliminary merged image with the \verb|merge_obs| script. We used \texttt{wavdetect} with a threshold $10^{-6}$ and wavelet scales $\sqrt{2},2,4,$ and 8 on the merged images in three bands ($0.5-2$, $2-7$, and $0.5-7$ keV) to generate a preliminary point source catalog.

\begin{deluxetable*}{ccccccccccchc}[ht]
\scriptsize
\tablecaption{\label{table:obslog}Log of \cxo\ observations.}
\tablehead{
\colhead{ObsID} & 
\colhead{R.A.\tablenotemark{a}} & 
\colhead{Dec.\tablenotemark{a}} & 
\colhead{Roll} & 
\colhead{Start Date} & 
\colhead{Exposure} & 
\colhead{X Shift} &
\colhead{Y Shift} &
\colhead{Detector} &
\colhead{Mode\tablenotemark{b}} & 
\colhead{Observer} & 
\nocolhead{Diff. Flag\tablenotemark{c}} &
\colhead{Bkg. Area}\\ 
\colhead{ } & 
\colhead{($\mathrm{{}^{\circ}}$)} & 
\colhead{($\mathrm{{}^{\circ}}$)} & 
\colhead{($\mathrm{{}^{\circ}}$)} & 
\colhead{} & 
\colhead{($\mathrm{ks}$)} & 
\colhead{(Sky pix.)} & 
\colhead{(Sky pix.)} &
\colhead{ } &
\colhead{ } & 
\colhead{ } & 
\colhead{ } &
\colhead{(arcmin$^2$)}}
\startdata
7863 & 184.708 & 14.415 & 57.5 & 2007-11-21 & 5.1 & 0.49 & -0.94 & ACIS-I & F & S. Mathur & Y & 25.2 \\
17462 & 184.735 & 14.447 & 89.3 & 2015-02-16 & 44.5 & 0.00 & 0.00 & ACIS-S & V & G. Garmire & Y & 24.9 \\
27362 & 184.663 & 14.429 & 194.2 & 2023-03-31 & 12.9 & 3.02 & -0.96 & ACIS-S & V & G. Garmire & N & 18.1 \\
27774 & 184.741 & 14.448 & 77.2 & 2024-01-08 & 23.3 & 0.31 & -0.14 & ACIS-S & V & G. Garmire & N & 19.8 \\
27775 & 184.662 & 14.428 & 195.4 & 2023-03-31 & 12.7 & 2.68 & -0.77 & ACIS-S & V & G. Garmire & N & 22.6
\enddata
\tablenotetext{a}{Aimpoint position.}
\tablenotetext{b}{Observing telemetry mode: F = `FAINT', VF = `VFAINT'.}
\tablecomments{The collected Chandra ObsIDs used in this work are available at the Chandra Data Archive (CDA) with DOI \dataset[10.25574/cdc.450]{\doi{https://doi.org/10.25574/cdc.450}}.}
\end{deluxetable*}

Point source properties were extracted using \acisx\ v2023Aug14 with our preliminary \texttt{wavdetect} source catalog as input. \acisx\ uses \texttt{MARX} (v5.5.3) to model the \cxo\ PSF at the position of each source, allowing optimal point-source masking for the study of the underlying diffuse emission. \acisx\ automatically generates event lists with events associated with point sources masked\footnote{Hereafter we refer to these as ``diffuse'' event lists rather than adopting the \acisx\ ``background'' nomenclature, to avoid confusion with the combined cosmic and instrumental background.}. We use these diffuse emission event lists as our primary science product for the analysis of the diffuse emission spectra in what follows.

\begin{figure*}
    \centering
    \includegraphics[width=0.925\textwidth]{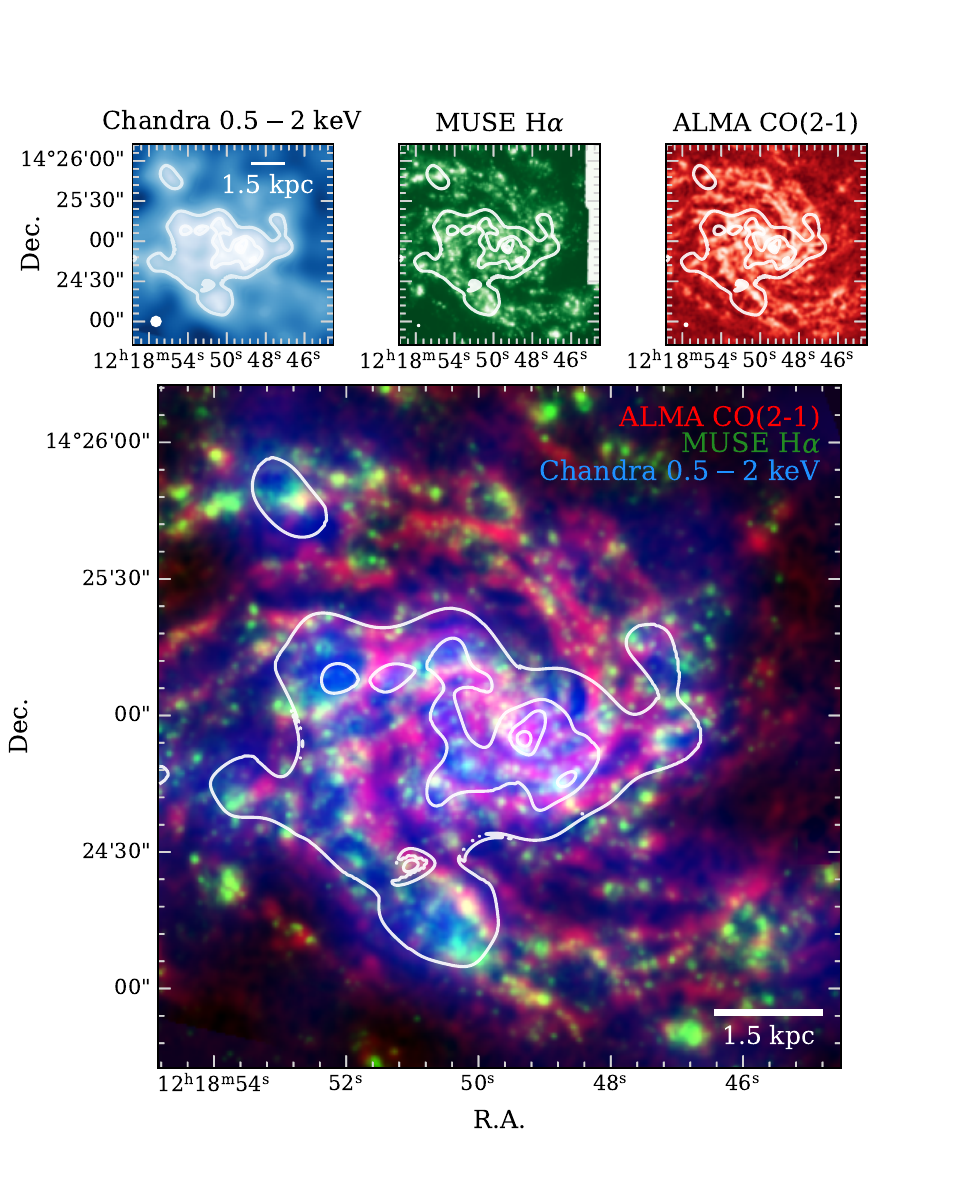}
    \caption{\label{fig:ISM_composite}Three phases of the ISM, imaged by Chandra, MUSE, and ALMA. \textit{Top Row}: From left to right, the adaptively smoothed soft band Chandra image created in \autoref{sec:data:xray}, the extinction-corrected MUSE $\rm H\alpha$ map, and the ALMA CO(2-1) moment 0 (integrated line flux) map using the ``broad'' masking \citep[see][]{leroy2021b}. Each image is displayed with an asinh stretch on a 99\% scale. The \added{white} circle in the lower left corner of the Chandra image represents the FWHM of the smoothing kernel near the center of the image; in the MUSE and ALMA images, the (nearly invisible) \added{white} circles represent the FWHM of the homogenized PSF and synthesized beam, respectively. \textit{Bottom:} Three-color composite of the images in the top row. \added{Several qualitative, kpc-scale trends are visible in this composite, though we caution that we have not performed an absolute astrometric reprojection of the Chandra image to the PHANGS frame and an $\approx 0\farcs24$ offset is expected (see \autoref{sec:data:xray})} In the spiral arms, the hot gas appears to fill large gaps near bright CO knots in the ALMA map \added{in at least on region around R.A.$=$12:18:52.2, Dec.$=+$14:25:10.0}, while associating with $\rm H\alpha$ emission on scales $\sim 1$ kpc. In both upper and lower panels, Chandra contours are plotted at arbitrary levels to guide the eye.}
\end{figure*}

For visualization purposes, we used \verb|merge_obs| to create merged images of the diffuse emission in the $0.5-2$ keV band and used \texttt{dmimgadapt} with minimum scale set to 5 sky pixels ($\approx 2\farcs5$) and maximum scale set to 15 sky pixels ($\approx 7\farcs5$) to smooth over the holes left by masking point sources. We require at least 16 counts under each smoothing kernel. Additionally, to estimate the contribution of the unresolved cosmic X-ray background and instrumental background to the diffuse emission image, we used the merged diffuse emission image to create a ``blank-sky'' image of the galaxy region by masking out the optical extent of the galaxy (defined \added{as the star formation rate surface density contour with \SFRD\ $= 10^{-2}~{\rm M_{\odot}~yr^{-1}~kpc^{-2}}$}, derived from the SFR map described in \autoref{sec:sfh}) and replacing events inside the galaxy region with events from a nearby background region with similar effective exposure using the \texttt{dmfilth} task with the \texttt{`DIST'} option. \added{The masked region was chosen ``by-eye'' to minimally encompass all of the clear diffuse X-ray emission while capturing the shape of the galaxy. The region is large enough such that the resulting blank-sky map is not highly sensitive to the shape or size of the masked region.} We smoothed the blank-sky image with the same global smoothing scale used for the diffuse emission map. The resulting soft-band diffuse emission map, with the blank-sky map subtracted, is shown in the top left panel of \autoref{fig:ISM_composite}.

\subsection{UV-to-IR Photometric Data} \label{sec:data:uvir}

We use the UV-to-IR photometric data cube produced by \citet{lehmer2024} following the procedures described in \citet{eufrasio2017}, including the bands listed in \autoref{table:bands}. \added{We direct interested readers to \citet{lehmer2024} for a more complete description of the UV-to-IR data reduction and preparation.} Briefly, the data were homogenized to common units, and bright foreground stars were identified and masked following the method of \citet{eufrasio2017} using adaptively sized circular masking regions as in \citet{lehmer2024}. The star-masked images were convolved to a common \psfsize\ PSF, a conservative choice that allows us to incorporate bands up to \herschel\ SPIRE $250~ \rm \mu m$, and projected to a common \pixsize\ pixel scale, which corresponds to a physical area scale \physsize\ at the distance of NGC 4254. 

\begin{figure*}
    \centering
    \includegraphics[width=0.95\textwidth]{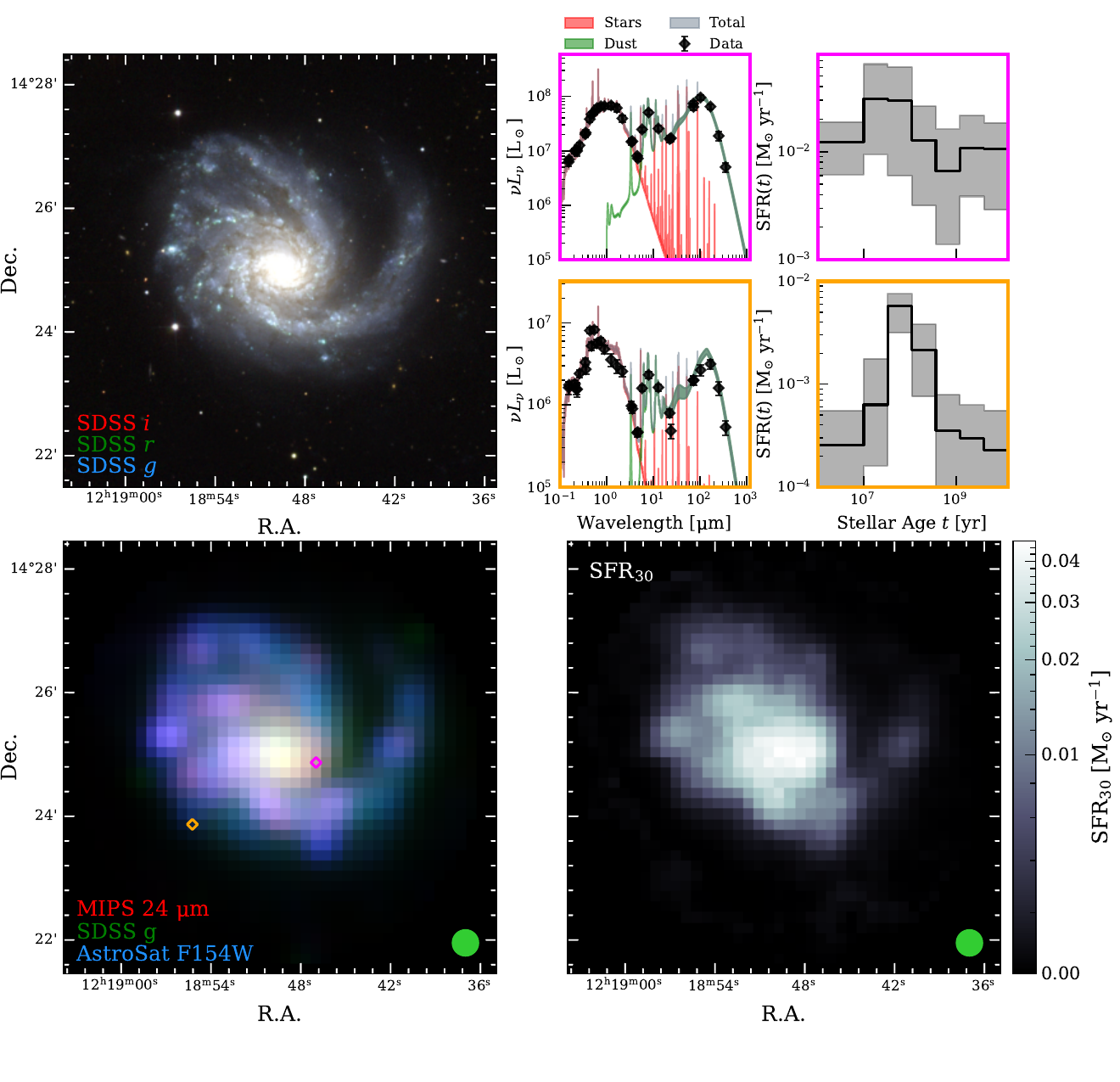}
    \caption{\label{fig:sidebyside} Schematic of the pixel-by-pixel UV-to-IR SED-fitting procedure. 
    \textit{Top left:} SDSS $gri$ composite of NGC 4254 with $0\farcs5$ pixels. NGC 4254 is a grand design spiral with prominent Northern and Southern arms and a faint third arm in the Northeast (all images are displayed with North up and East to the left).
    \textit{Bottom left:} A three-color composite extracted from the PSF-matched multiwavelength data cube, with red = \spitzer\ MIPS $24~\rm \mu m$, green = SDSS $g-$band, and blue = \astrosat\ F154W with two example pixels marked by color-coded diamonds. The SED fits and SFHs for these same example pixels are shown in the upper-right panel. The pixel size \added{in this panel} is $10''$, 20 times larger than the SDSS image.
    \textit{Upper right:} We show the full UV-optical-IR SED fits and the corresponding piecewise SFH for the \added{two pixels marked in the bottom left panel}; \added{the border colors of the plots correspond to the color of the two diamond markers in that panel}. Note the different axis scales for both SED and SFH plots. The pixel in the inner Northern arm \added{(magenta border; magenta diamond in the lower left panel)} has significant ongoing star formation and is both significantly brighter and more attenuated; the pixel in an inter-arm region in the outer disk \added{(orange border; orange marker in lower left panel)} is fainter and last experienced significant star formation $\gtrsim30$ Myr ago.
    \textit{Bottom right:} The $0-30$ Myr average SFR ($\rm SFR_{30}$) map derived from the full set of pixel-by-pixel SED fits. In the bottom left and bottom right panels, the green circle shows the $25''$ FWHM of the homogenized PSF.}
\end{figure*}

The background was measured in an annulus centered on the galaxy with inner and outer ellipses constructed by scaling the $K-$band extent of the galaxy (an ellipse defined by major and minor radii of $a=1.7$ arcmin and $b=1.62$ arcmin, and a position angle of 23.5 degrees) by factors of 2 and 2.3, respectively. \added{The background annulus scales were manually chosen by inspection of the image; the resulting photometry is not highly sensitive to the size of the annulus.} The background regions for each band typically contain $\approx 400$ pixels, depending on the field of view of the observation. We show a multiwavelength composite extracted from the datacube in \added{the lower left panel of} \autoref{fig:sidebyside}.

Each bandpass in the data cube was corrected for Galactic extinction using the \citet{fitzpatrick1999} extinction curve, assuming $A_V = 0.1034$ mag, as retrieved from the IRSA DUST web application\footnote{\url{https://irsa.ipac.caltech.edu/applications/DUST/}} using the \citet{schlafly2011} recalibration.

\begin{deluxetable}{lcc}
    \scriptsize
    \tablecaption{\label{table:bands}Bands used for UV-IR SED fitting.}
    \tablehead{
    \colhead{Observatory/Instr.} & 
    \colhead{Bands} & 
    \colhead{$\sigma^{\rm cal}_{f}$\tablenotemark{a}}
    }
    \startdata
    \textit{GALEX} & FUV, NUV & 0.15 \\
    \textit{AstroSat}/UVIT & F154W & 0.05 \\
    \textit{Swift}/UVOT & UVW2, UVM2, $B$, $V$ & 0.05 \\
    SDSS & $u,g,r,i,z$ &  0.05 \\
    2MASS & $J,H,K_s$ & 0.10 \\
    \textit{WISE} & W1, W2, W3, W4 & 0.07 \\
    \textit{Spitzer}/IRAC & Ch1, Ch2, Ch3, Ch4 & 0.05 \\
    \textit{Spitzer}/MIPS & Ch1 & 0.05\\
    \textit{Herschel}/PACS & P1, P2, P3 & 0.05  \\
    \textit{Herschel}/SPIRE & S1, S2 & 0.15 \\
    \enddata
\tablenotetext{a}{Uncertainty in the flux calibration, given here as a fraction of the flux. For instrument/filter combinations where flux calibration uncertainties are not readily available, we conservatively assume $5\%$.}
\end{deluxetable}

Uncertainties in the pixel fluxes were calculated based on the background and the flux calibration uncertainties given in \autoref{table:bands}, such that the total variance of the flux density for a pixel, $\sigma_{f_\nu}^2$, is

\begin{equation}
    \sigma_{f_\nu}^2 = \left(1+1/M \right) (\sigma^{\rm bkg}_{f_\nu})^2 + (\sigma^{\rm cal}_{\nu} f_\nu)^2,
\end{equation}
where $(\sigma^{\rm bkg}_{f_\nu})^2$ is the variance of the flux density in the large background region, $M$ is the number of pixels in the background region, and $\sigma^{\rm cal}_{\nu}$ is the fractional calibration uncertainty. At the chosen pixel scales, the uncertainties on the flux are dominated by the calibration uncertainty, \added{exceeding the background uncertainty by a factor of $\sim 10$ for the brightest pixels}. \added{The background uncertainty becomes more significant in fainter regions, with the ratio between calibration and background uncertainties approaching unity in the fainter interarm regions. At the outskirts of the galaxy, the background uncertainty becomes the dominant contribution by a factor of $\sim10-$100.}

\subsection{Ancillary PHANGS Data and Products}\label{sec:data:phangs}

As part of the PHANGS project, NGC 4254 has ALMA CO(2-1) line imaging, targeting the entire star-forming region of the galaxy \added{(dataset ID 2015.1.00956.S; PI: Leroy)}. \added{The PHANGS-ALMA survey design is discussed in \citet{leroy2021a}, and the ALMA data reduction is described by \citet{leroy2021b}. The ALMA observations have been processed to combine interferometric and single-dish observations.} \added{The PHANGS-MUSE survey provides IFU spectral imaging from the Multi-Unit Spectrographic Explorer on the Very Large Telescope, again targeting the star-forming central region of the galaxy (dataset ID 1100.B-0651; PI: Schinnerer). The PHANGS-MUSE survey and data reduction process are described in \citep{emsellem2022}.} \added{We also use} archival 21 cm imaging from the Very Large Array \citep{phookun1993, chung2009}, which has been reprocessed and aggregated into high-level measurement tables by \citet{sun2023}. These ancillary observations allow us to place the hot phase of the ISM traced by our Chandra maps in context with the cold phase and with the gas ionized by massive stars. 

We use the $2''$ beam ALMA maps from PHANGS-ALMA v4.0\footnote{\url{https://www.canfar.net/storage/vault/list/phangs/RELEASES/PHANGS-ALMA}} \added{described by \citet{leroy2021b}} and the ``convolved and optimized'' (``copt'') line maps from the PHANGS-MUSE DR1.0\footnote{\url{https://www.canfar.net/storage/vault/list/phangs/RELEASES/PHANGS-MUSE/DR1.0}} \added{described by \citet{emsellem2022}}. The copt maps are PSF-homogenized to the single largest PSF of any of the observations making up the map, in this case a FWHM of $0\farcs89$. We have extinction-corrected the MUSE line maps using the observed $\rm H\alpha/H\beta$ ratio maps and the \citet{odonnell1994} curve with $R=3.1$, assuming a theoretical $\rm H\alpha/H\beta = 2.86$ \added{\citep[following, e.g.,][]{groves2023}}. The ALMA moment 0 (line intensity) and MUSE $\rm H\alpha$ maps are shown in the top right and top middle panels of \autoref{fig:ISM_composite}, respectively.

\citet{sun2022} divided the PHANGS galaxies, including NGC 4254, into regular hexagonal tilings and calculated a variety of physical quantities, including multi-scale molecular densities, averaged across each tile. The tiling is centered on the optical center of the galaxy, and the hexagons have an inscribed physical diameter of 1.5 kpc ($\approx 23\farcs 6$ for NGC 4254). We use v4.2 of these high-level measurement tables (\added{also called ``megatables''}), which are described in \added{Appendix C of \citet{sun2022} and Appendix A} of \citet{sun2023}. The molecular gas surface density is calculated from ALMA CO data as
\begin{equation}
    \Sigma_{\rm mol} = \alpha_{\rm CO(1-0)}R_{21}^{-1}~I_{\rm CO(2-1)} \cos i, 
\end{equation}
where $I_{\rm CO(2-1)}$ is the CO\,(2--1) line integrated intensity (i.e., moment 0), $R_{21} = 0.65$ is the $\rm CO(2{-}1)/CO(1{-}0)$ line ratio, $\alpha_{\rm CO(1-0)}$ is a metallicity-dependent CO-to-H$_2$ conversion factor following the prescription used by \citet{sun2020}, and $\cos i$ is an inclination correction factor. The catalogs provide the area-weighted surface brightness calculated over each hexagon, which we denote $\Sigma^{1.5~\rm kpc}_{\rm mol}$, as well and the mass-weighted average (over each hexagon) of the surface mass density of individual 150 pc-scale molecular cloud complexes, which we denote $\langle \Sigma^{\rm 150~pc}_{\rm mol}\rangle$. These measurement tables also include estimates of the cloud-scale turbulent molecular pressure, which supports the molecular clouds against external pressures and self gravitation, and the dynamical equilibrium pressure $P_{\rm DE}$, which represents the required pressure for the bulk ISM to stay in equilibrium in the gravitational potential of the galaxy.
We \added{summarize the quantities above in \autoref{tab:molecularprops} and} refer interested readers to \citet{sun2020}, \citet{sun2022}, and \citet{sun2023} for further details.

\begin{deluxetable*}{l c c c c}
    \tablecaption{\label{tab:molecularprops}Summary of the quantities we adopt from the PHANGS measurement tables.}
    \tablehead{
        \colhead{Symbol} & \colhead{Definition} & \colhead{Phys. Scale} & \colhead{Column Name\tablenotemark{a}} & \colhead{Reference}
    }
    \startdata
    $\Sigma^{1.5~\rm kpc}_{\rm mol}$ & Area-weighted molecular gas surface density & 1.5 kpc & \texttt{Sigma\_mol} & \citet{sun2022}\\
    $\langle \Sigma^{\rm 150~pc}_{\rm mol}\rangle$ & Mass-weighted average molecular gas surface density of molecular clouds & 150 pc & \texttt{<Sigma\_mol\_obj\_150pc>} &  \citet{sun2022}\\
    $P_{\rm mol}$ & Flux-weighted average internal turbulent pressure of molecular clouds & 150 pc & \texttt{<P\_turb\_pix\_150pc>} & \citet{sun2020}\\
    $P^{1.5~\rm kpc}_{\rm DE}$ & Kpc-scale dynamical equilibrium pressure & 1.5 kpc & \texttt{P\_DE} & \citet{sun2020} \\
    $P^{150~\rm pc}_{\rm DE}$ & Flux-weighted average dynamical equilibrium pressure of molecular and atomic clouds & 150 pc & \texttt{<P\_turb\_pix\_150pc>} & \citet{sun2020} \\
    \enddata
    \tablenotetext{a}{Column name in v4.2 of the PHANGS measurement tables.}
\end{deluxetable*}

\citet{groves2023} presented the PHANGS-MUSE nebular catalog, which contains candidate H II regions identified and classified in the PHANGS galaxies using MUSE data. We adopt the metallicities derived by \citet{groves2023} using the S calibration method (see their section 4.3 for details). Specifically, since our analyses mostly proceed at scales larger than a typical H II region, we use the radial metallicity gradient fit by \citet{groves2023}:
\begin{equation}
    12 + \log{\rm O/H} = 8.59 - 0.028 \frac{r}{r_{\rm eff}},
\end{equation}
where $r_{\rm eff} = 0.6$ arcmin and the scatter around this relation is estimated at $\sigma = 0.03$ dex. To convert between Oxygen abundance and heavy-element mass fraction, we assume the Solar gas phase oxygen abundance to be $12 + \log{\rm O/H} = 8.69$ \citep{asplund2009}, which we take to correspond to a heavy-element mass fraction $Z_{\odot} = 0.02$. 

For visualization purposes, we also adopt the definition of the spiral arms from the ``environment masks'' generated from Spitzer 3.6 \micron\ data by \citet{querejeta2021}. \added{The arm masks were defined by fitting logarithmic spiral functions to 3.6 \micron\ images of the PHANGS galaxies, and consequently trace stellar structures. We plot the outline of the ``arm'' environment category in \autoref{fig:hexplot}.}

\section{Star Formation History Maps} \label{sec:sfh}

Linking the X-ray emission to SNe (and hence star formation) across the galaxy requires estimates of the CCSNe rates. The CCSNe rates peak at times later (at stellar ages $\sim 10-20$ Myr) than those probed by the often-used H$\alpha$-derived SFR. We thus elect instead to derive star formation rates by SED fitting with BPASS stellar population synthesis models, which has the additional advantage that star formation rates for sub-galactic regions can be directly translated to supernova rates using the CCSNe rates from BPASS. We use the BPASS v2.2.1 stellar population models, with a \citet{chabrier2003} \added{initial mass function (IMF)}.

To estimate the spatially-resolved star formation rate of the galaxy, we performed pixel-by-pixel SED fitting to the UV-to-IR data cube described in \autoref{sec:data:uvir}, using the newly-released Python version of the \lgh\ SED-fitting code\footnote{The source code is available at \url{https://www.github.com/ebmonson/lightningpy} and the documentation can be found at \url{https://lightningpy.readthedocs.io}. \added{A frozen copy of the version v2025.1.0 used for this work is available with DOI \dataset[10.5281/zenodo.18011894]{\doi{10.5281/zenodo.18011894}}}.} This method has previously been used to derive robust star formation rates across M51 \citep{eufrasio2017, lehmer2017} and has the advantage of not assuming a fixed SFH shape, unlike commonly-adopted SFR scaling relations (see also \autoref{fig:sfrd_comparison}). Additionally, our use of IR data allows us to self-consistently account for obscured star formation by assuming energy balance between the attenuated UV-optical luminosity and reprocessed mid- and far-IR luminosity. 

We have updated \lgh\ to include new nebular models based on \cloudy\ simulations, which allow more flexibility in the nebular emission associated with the youngest stellar populations. The \cloudy\ simulations are produced assuming an open geometry, with constant pressure. For this work, we use the simulation grids that exclude the impact of dust grains on the nebular component, and we assume $n_{\rm H} = 100~{\rm cm^{-3}}$. \added{Since we do not directly fit nebular lines, our SED fits are insensitive both to the choice of density for the nebular component and to our exclusion of dust grain effects on the nebular component. The primary contribution of the nebular models is thus a weak continuum component, which we include for completeness.}

In contrast to earlier versions of \lgh, the new version allows the metallicity of the stellar population to vary (and the metallicity of the nebula, which is fixed to the stellar metallicity in the current version of the model), allowing us to account for variations in metallicity across the galaxy. We assign each pixel in the data cube a metallicity based on the \citet{groves2023} metallicity gradient, using the distance from the pixel center to the galactic center (we adopt R.A. = \galctrra\ deg, Dec. = \galctrdec\ deg). The metallicity predicted by the gradient is used as the mean of a normal prior distribution on the stellar metallicity, with $\sigma=0.03$ dex, the estimated scatter around the relation. We note that the stellar and ISM metallicities in the current version of \lgh\ are tied together. The ISM metallicity, however, is comparable only to the enrichment of the youngest generations of stars: when considering contributions from older stars, the stellar metallicity gradients in galaxies tend to be steeper than gas-phase metallicity gradients, with the light-averaged stellar metallicity less enriched by 0.3-0.5 dex compared to the ISM for galaxies in similar mass ranges to NGC 4254 \citep{lian2018}. Our broadband photometric SED fitting and the results we present should not be strongly affected by the offset between old stellar metallicity and ISM metallicity, given our focus on the young stars whose SNe drive the diffuse X-ray emission.

For each pixel in our multiwavelength data cube, we fit the SED assuming a piecewise-constant SFH with \nbins\ stellar age bins, where the first bin spans $0-10$ Myr, and the remainder are log-spaced from 10 Myr $-$ 13.4 Gyr. We adopt the ``modified Calzetti'' attenuation curve as implemented in \lgh, which uses the variable UV slope and 2175 \AA\ Drude-profile bump introduced by \citet{noll2009}. Dust emission is modeled with the ``restricted'' \citet[][DL07]{draine2007} model, with $\alpha=2$ and $U_{\rm max} = 3\times10^5$ \added{(respectively the power law index and maximum intensity describing the distribution of stellar radiation field intensities)}. We assumed energy balance between the attenuated luminosity from the stellar population and the total IR luminosity of the dust model, such that the normalization of the \citetalias{draine2007} model is not a free parameter of the model. \added{Interested readers are referred to \citet{doore2023} for a more complete outline of the dust emission model implementation in \texttt{Lightning} and \texttt{lightning.py}.}

We summarize the model specification and the priors we adopt on each parameter in \autoref{table:lghspec}. \added{We default to loose, uninformative priors on most parameters. We allow the star formation coefficients and $V-$band optical depth to vary well above the expected values for single pixels in order to capture the large uncertainties on per-pixel star formation rate. The $\delta$ parameter of the attenuation curve and the $U_{\rm min}$, $\gamma$, and $q_{\rm PAH}$ parameters of the dust emission model were allowed to vary uniformly over their full defined range. 

We only adopt normal-distribution priors on two parameters: $Z$, as described above, and $\log U_{\rm ion}$, the dimensionless ionization parameter. Since we fit no nebular lines, the results are insensitive to the choice of the prior for the ionization parameter. We chose a broad normal prior for $\log U_{\rm ion}$, centered at -2.5, effectively allowing $\log U_{\rm ion}$ to vary over its full range while slightly penalizing extreme values. The ionization parameter for typical H II regions in this galaxy is -1.7 \citep{groves2023}, larger than our prior mean; however our $10''$ pixels are significantly larger than individual H II regions and we would thus expect any nebular signatures in our data to be diluted.} 

We sampled the posteriors using the \texttt{emcee} affine-invariant MCMC sampler, using an ensemble of \nwalkers\ walkers and running them for \nsteps\ steps. We adopted an additional 10\% model uncertainty term for fitting, added in quadrature to the data uncertainties. To construct the final chain, we discard the first \ndiscard\ steps for each walker and retain every \nthin\ samples. We check for independence of the samples in the resulting chains by computing the autocorrelation time; we find autocorrelation times $\approx 1-2$ steps, indicating the samples in our chains are independent. Fit quality, as measured by the stacked residuals, is good across the map: our SED models show no systematic residuals across the 1203 pixels fit (see \autoref{sec:quality}). \added{Representative SED fits and the corresponding SFH models are shown in the upper right panel of \autoref{fig:sidebyside}.} 

The supernova rate per stellar mass peaks between $10-30$ Myr in our stellar population models, declining thereafter. To probe the youngest component of the stellar population, which produces the massive stars whose supernovae drive shocks into the ISM, we calculated the SFR by averaging over the first two bins in our SFH, yielding the average SFR from 0 to $\approx 30$ Myr (in what follows we refer to this as ${\rm SFR_{30}}$ for brevity). We show the ${\rm SFR_{30}}$ map in the lower right panel of \autoref{fig:sidebyside}. 

As a consistency check, we compare our SED fitting results and SFR to existing measurements from the PHANGS project in \autoref{sec:quality}. \added{The 0-10 Myr SFR and the short-timescale SFR indicators collated in the \citet{sun2023} measurement tables (see left panel of \autoref{fig:sfrd_comparison}) agree within 0.2 dex, with the exception of the ``re-calibrated'' SFR indicators at the highest \SFRD, which are around 0.3 dex smaller than our estimate of the 0-10 Myr \SFRD. We see that our 0-30 Myr \SFRD\ measurement is systematically higher than the one- and two-point SFR indicators by up to 0.4 dex (see right panel of \autoref{fig:sfrd_comparison}), due to our assumption of a flexible SFH model: the SFH in most regions peaks at ages older than 10 Myr, such that ${\rm SFR_{30}}$ is larger than the 0-10 Myr SFR. We also check for consistency between our dust SED parameters and those measured by \citet{chastenet2025} (see \autoref{fig:dust_comparison}). While our implementations of the \citet{draine2007} dust emission model are slightly different, we find that our measurements are very similar, typically varying by only 0.1 dex in the $U_{\rm min}$ parameter.} 


\begin{deluxetable*}{l c c c}
    \tablecaption{\label{table:lghspec}UV-IR SED model specification in \lgh.}
    \tablehead{\colhead{Model Component} & \colhead{\lgh\ Name} & \colhead{Parameter} & \colhead{Prior} \\
               \colhead{(1)}               & \colhead{(2)}          & \colhead{(3)}         & \colhead{(4)}}
    \startdata
    SFH                       & \texttt{Piecewise-Constant}  & $\{\psi_i\}$          & $\mathcal{U}(0,1)$ \\
    \midrule
    Attenuation Curve         & \texttt{Modified-Calzetti}   & $\tau_{V, \rm diff}$ & $\mathcal{U}(0,3)$ \\
                              &                              & $\delta$             & $\mathcal{U}($-2.3, 0.4$)$ \\
                              &                              & $\tau_{V, \rm BC}$   & 0 \\
    \midrule
    Stellar Population \& Nebula & \texttt{BPASS-A24}        & $Z$                  & $\mathcal{N}(Z(r)$\tablenotemark{a}, 0.03 dex) \\
                                 &                           & $\log U_{\rm ion}$             & $\mathcal{N}(-2.5,0.75)$ \\
    \midrule
    Dust Emission                & \texttt{DL07}              & $\alpha$             & 2 \\
                                 &                           & $U_{\rm min}$        & $\mathcal{U}(0.1, 25)$ \\
                                 &                           & $U_{\rm max}$        & $3\times10^{5}$\\
                                 &                           & $\gamma$             & $\mathcal{U}(0,1)$\\
                                 &                           & $q_{\rm PAH}$         & $\mathcal{U}(0.0047, 0.0458)$\\
    \enddata
    \tablecomments{1: Component of the SED model; 2: Model choice in \lgh; 3: Model parameter; 4: Corresponding prior, where $\mathcal{U}(a,b)$ is the Uniform distribution on the interval $[a,b)$, and $\mathcal{N}(\mu,\sigma)$ is the normal distribution with mean $\mu$ and standard deviation $\sigma$. Parameters with a single value are fixed.}
    \tablenotetext{a}{Where $Z(r)$ is the metallicity appropriate for the pixel's distance from the galaxy center, assuming the metallicity gradient estimated by \citet{groves2023}.}
\end{deluxetable*}

\section{Bayesian X-ray Spectral Fitting} \label{sec:spec}

We use the 1.5 kpc hexagonal tiling from \citet{sun2022} as a convenient way to define spectral extraction regions for the diffuse X-ray emission in which the properties of the molecular and atomic ISM are already well-known. We evaluated the ${\rm SFR_{30}}$ surface density 
\begin{equation}
    \SFRD = ({\rm SFR_{30}} / A) \cos i,
\end{equation}
within each hexagonal tile, where $A$ is the projected area of the tile (1.95 kpc$^2$) and $i$ is the inclination of the galaxy (34.4 deg). We grouped tiles in five \SFRD\ bins to accumulate enough counts to measure hot gas temperatures and luminosities: 
$0.01-0.03$, $0.03-0.05$, $0.05-0.07$,  $0.07-0.09$, and $0.09-0.13~{\rm M_{\odot}~yr^{-1}~kpc^{-2}}$.
The soft-band ($0.5-2$ keV) and full-band ($0.5-7$ keV) net (i.e., background-subtracted) counts are given for each \SFRD\ bin in \autoref{tab:stackparams}. \added{Our choice to bin on \SFRD\ rather than other sub-galactic properties is motivated by prior studies connecting the galaxy-integrated X-ray properties to SFR and the observed tight X-ray luminosity--SFR correlation \citep[e.g.][]{mineo2012}. Implicitly, this choice assumes that \SFRD\ is the dominant parameter controlling sub-galactic variations in the X-ray surface brightness and hot gas temperature across the galaxy. While the metallicity of the gas may also play a role in sub-galactic variations, as higher-metallicity gas can cool more efficiently at fixed temperature (producing larger luminosities at fixed temperature and density), we note that the radial metallicity gradient of this galaxy is shallow \citep{groves2023} and the metallicity varies little across the disk. Our choice of bins also roughly corresponds to a radial binning: given the radial decline in \SFRD, the lowest-\SFRD\ bins are the outer disk, and the higher-\SFRD\ bins encompass the inner disk.}

We used the \texttt{specextract} \texttt{CIAO} script to extract source spectra, background spectra, auxiliary response files (ARFs), and response matrix files (RMFs) for each group of tiles from the diffuse event lists for each ObsID. For each ObsID, we adopted a large background region on the same chip as the galaxy; the sizes of the background regions are given in \autoref{table:obslog}. We treat the grouped cells as stacks in what follows to estimate the average parameters of the diffuse emission on 1.5 kpc scales. Since the low-\SFRD\ bins naturally contain more tiles, this treatment ensures that we can fit spectra with $>100$ net counts per \SFRD\ bin, even in the low-\SFRD\ outskirts of the galaxy.

\begin{deluxetable*}{cccccccccccc}
\scriptsize
\tablecaption{\label{tab:stackparams}Sampled quantities for each of the \SFRD\ bins and the X-ray extent of the galaxy, assuming a single-temperature plasma model.}

\tablehead{
\multicolumn{2}{c}{$\Sigma_{\rm SFR}$} & 
\colhead{} & \colhead{} &
\colhead{} & \colhead{} & 
\multicolumn{3}{c}{\texttt{apec} Luminosity\tablenotemark{a}} & 
\colhead{} & \colhead{} \\
\multicolumn{2}{c}{$(\mathrm{M_{\odot}\,yr^{-1}\,kpc^{-2}})$} &
\multicolumn{2}{c}{Net Cts.} & 
\colhead{$kT$} & 
\colhead{$\log N_H$} & 
\multicolumn{3}{c}{$(\mathrm{10^{38}\,erg\,s^{-1}})$} &
\colhead{$n_e\sqrt{f}$} & 
\colhead{$n_e^2 V\sqrt{f}$} &
\colhead{$P\sqrt{f}$}\\
\cmidrule(lr){1-2} \cmidrule(lr){3-4} \cmidrule(lr){7-9}
\colhead{Low} & 
\colhead{High} & 
\colhead{0.5$-$2 keV} & 
\colhead{0.5$-$7 keV} & 
\colhead{$(\mathrm{keV})$} & 
\colhead{$(\mathrm{cm^{-2}})$} & 
\colhead{0.5$-$2 keV, abs.\tablenotemark{b}} & 
\colhead{0.5$-$2 keV, intr.\tablenotemark{c}} & 
\colhead{0.3$-$10 keV, intr.} &
\colhead{$(\mathrm{0.01\,cm^{-3}})$} & 
\colhead{$(\mathrm{10^{-3}\,kpc^{3}\,cm^{-6}}$)} &
\colhead{$(10^5\,k_B\,\mathrm{K\,cm^{-3}})$}
}
\startdata
0.01 & 0.03 & 346.7 & 564.9 & $0.17_{-0.02}^{+0.06}$ & $21.54_{-0.54}^{+0.13}$ & $0.37_{-0.04}^{+0.04}$ & $1.94_{-1.34}^{+1.48}$ & $5.79_{-4.42}^{+6.15}$ & $7.13_{-4.08}^{+4.24}$ & $2.40_{-1.96}^{+3.70}$ & $2.70_{-1.13}^{+1.07}$ \\
0.03 & 0.05 & 348.9 & 416.9 & $0.32_{-0.04}^{+0.05}$ & $20.59_{-0.38}^{+0.43}$ & $1.09_{-0.09}^{+0.10}$ & $1.41_{-0.17}^{+0.37}$ & $2.53_{-0.38}^{+0.92}$ & $3.80_{-0.44}^{+0.79}$ & $0.68_{-0.15}^{+0.32}$ & $2.76_{-0.19}^{+0.22}$ \\
0.05 & 0.07 & 268.8 & 319.5 & $0.27_{-0.06}^{+0.05}$ & $21.06_{-0.61}^{+0.38}$ & $1.66_{-0.16}^{+0.18}$ & $2.77_{-0.71}^{+2.77}$ & $5.68_{-1.92}^{+7.94}$ & $5.93_{-1.25}^{+3.74}$ & $1.66_{-0.63}^{+2.76}$ & $3.59_{-0.37}^{+1.00}$ \\
0.07 & 0.09 & 128.3 & 156.9 & $0.55_{-0.18}^{+0.15}$ & $20.59_{-0.41}^{+0.55}$ & $1.74_{-0.24}^{+0.25}$ & $2.20_{-0.33}^{+0.78}$ & $3.33_{-0.52}^{+1.58}$ & $3.70_{-0.37}^{+1.28}$ & $0.65_{-0.12}^{+0.52}$ & $4.66_{-0.75}^{+0.79}$ \\
0.09 & 0.13 & 174.3 & 199.0 & $0.27_{-0.06}^{+0.05}$ & $20.73_{-0.50}^{+0.59}$ & $3.58_{-0.43}^{+0.39}$ & $4.95_{-0.85}^{+4.43}$ & $9.77_{-2.34}^{+13.07}$ & $7.79_{-1.26}^{+4.79}$ & $2.87_{-0.85}^{+4.61}$ & $4.82_{-0.38}^{+1.18}$ \\
\midrule
\multicolumn{12}{c}{Galaxy-Integrated Properties\tablenotemark{d}} \\
\midrule
\multicolumn{2}{c}{$0.045 \pm 0.001$} & 972.8 & 1186 & $0.19_{-0.04}^{+0.06}$ & $21.50_{-0.40}^{+0.20}$ & $36.3_{-2.1}^{+1.8}$ & $152_{-87}^{+234}$ & $411_{-273}^{+875}$ & $11.0_{-5.1}^{+11.9}$ & $40.6_{-29.1}^{+136}$ & $4.60_{-1.36}^{+3.05}$
\enddata
\tablenotetext{a}{Luminosities are normalized by the number of tiles in the bin for ease of comparison.}
\tablenotetext{b}{Without absorption correction.}
\tablenotetext{c}{With absorption correction.}
\tablenotetext{d}{Properties derived from the spectrum extracted within the X-ray extent of the galaxy; see \autoref{sec:galint}. In lieu of the upper and lower \SFRD\ bounds reported for the \SFRD\ bins, we report the \SFRD\ within the X-ray extent.}
\end{deluxetable*}

Given the relatively small number of counts even after stacking, we adopted a Bayesian X-ray spectral fitting framework using \texttt{BXA} \citep{buchner2014}, which uses the \texttt{UltraNest} nested sampler \citep{buchner2021} to perform Bayesian inference with X-ray spectral models. We used v4.16.0 of the \texttt{Sherpa} X-ray spectral modeling code \citep{sherpa2001, sherpa2007} to define spectral models. We chose a source spectral model incorporating power law emission from un-subtracted, un-resolved X-ray binaries, an \texttt{apec} component representing a single temperature plasma in collisional ionization equilibrium (CIE), and both intrinsic and Galactic absorption modeled with two independent \texttt{tbabs} components. The intrinsic absorption model was applied to both the X-ray binary and plasma models with the same column density\footnote{In \texttt{XSpec} terms, the model is \texttt{tbabs * tbabs * (apec + pow)}}. We retrieved the Galactic neutral Hydrogen column density $N^{\rm gal}_H = 2.7 \times 10^{20}~{\rm cm^{-2}}$ along the line of sight to the galaxy using the \texttt{CIAO} \texttt{Colden} and the \citet{dickey1990} dataset.
We calculated an average metallicity for each of the \added{\SFRD\ bins} using the \citet{groves2023} metallicity gradient. These metallicities are given in \autoref{tab:NHprior}. The ejecta from supernovae will be enriched compared to the average ISM metallicity; as such, the average metallicity may be an underestimate in regions with significant spectral contributions from reverse-shocked ejecta in unresolved SNe remnants, \added{which may result in an overestimate of the luminosity of our single fixed-abundance \texttt{apec} model}. However, we expect the majority of the emission to be due to ISM material swept up and shocked by outflowing SNe winds (see \autoref{sec:wind}) and thus adopt the ISM metallicity in the absence of enough signal-to-noise to fit variable-abundance plasma models. We fixed the photon index of the power law component to $\Gamma = 1.8$, a typical assumption for XRBs \added{\citep[e.g.][]{swartz2004}}. 

The background spectral model is a power law superimposed with 8 Gaussians: two Gaussians model the soft rise in the background, while the remaining 6 model instrumental lines. This is similar to the empirical \cxo\ background model included in \texttt{BXA} \citep[see][Appendix A]{buchner2014}, where we use a power law for the underlying continuum rather than a constant. The shape and level of the background varies per observation, and thus we fit the background model for each ObsID independently. 

We set uninformative priors on the log-scaled normalizations of the power law and \texttt{apec} models and a uniform prior from $0.1-1$ keV on $kT$. To anchor our estimates of the intrinsic column density, we adopt a prior on $\log N_{\rm H}$ based on the H I surface densities reported in v4.2 of the PHANGS megatables \citep{sun2023}. We calculate the average column density for the hexagonal tiles in each \SFRD\ bin, assuming that half the atomic gas lies along the line of sight (i.e., above the X-ray emitting gas), and adopt the standard deviation of the tiles as the standard deviation of a normal prior. We list the prior mean and standard deviation in \autoref{tab:NHprior}. The large physical areas covered by most of our \SFRD\ bins are such that the priors are wide, with $\sigma$ up to a factor of ten. 
We note that this prior and the definition of our spectral model together imply the assumption of a disk-like geometry for the X-ray emitting plasma, in which the plasma is coincident with the neutral atomic Hydrogen disk. 
Gas sitting well above the galactic plane in a spherical geometry in the halo would have a correspondingly lower column density, decoupled from the density of the H I disk, due to the smaller amount of cold gas along the line of sight.
Additionally, our model applies the same $N_{\rm H}$ to the power law model which represents non-detected XRBs, though we would normally expect XRBs to be more obscured than the diffuse, less-embedded plasma. Since we mask the majority of emission from bright XRBs, we do not expect this assumption to strongly affect the results.

\begin{deluxetable}{ccccc}
\tablecaption{\label{tab:NHprior}Normal prior on $\log N_{H}$ and assumed metallicities, based on the H I surface density from 21 cm observations and the  metallicity gradient from PHANGS-MUSE, respectively.}
\tablehead{
\multicolumn{2}{c}{\SFRD} &
\multicolumn{2}{c}{$\log N_H$} &
\colhead{$12 + \log {\rm O/H}$} \\
\multicolumn{2}{c}{($\rm M_{\odot}~yr^{-1}~kpc^{-2}$)}&
\multicolumn{2}{c}{($\rm cm^{-2}$)}&
\nocolhead{}\\
\cmidrule(lr){1-2}
\cmidrule(lr){3-4}
\colhead{Low}&
\colhead{High}&
\colhead{Mean}&
\colhead{Std}&
\nocolhead{}
}
\startdata
0.01 & 0.03 & 20.8 & 0.72 & 8.52\\
0.03 & 0.05 & 20.8 & 0.86 & 8.55\\
0.05 & 0.07 & 20.8 & 1.00 & 8.56\\
0.07 & 0.09 & 20.7 & 0.93 & 8.57\\
0.09 & 0.13 & 20.7 & 1.19 & 8.58\\
\enddata
\end{deluxetable}

\begin{figure*}
    \centering
    \includegraphics[width=0.85\textwidth]{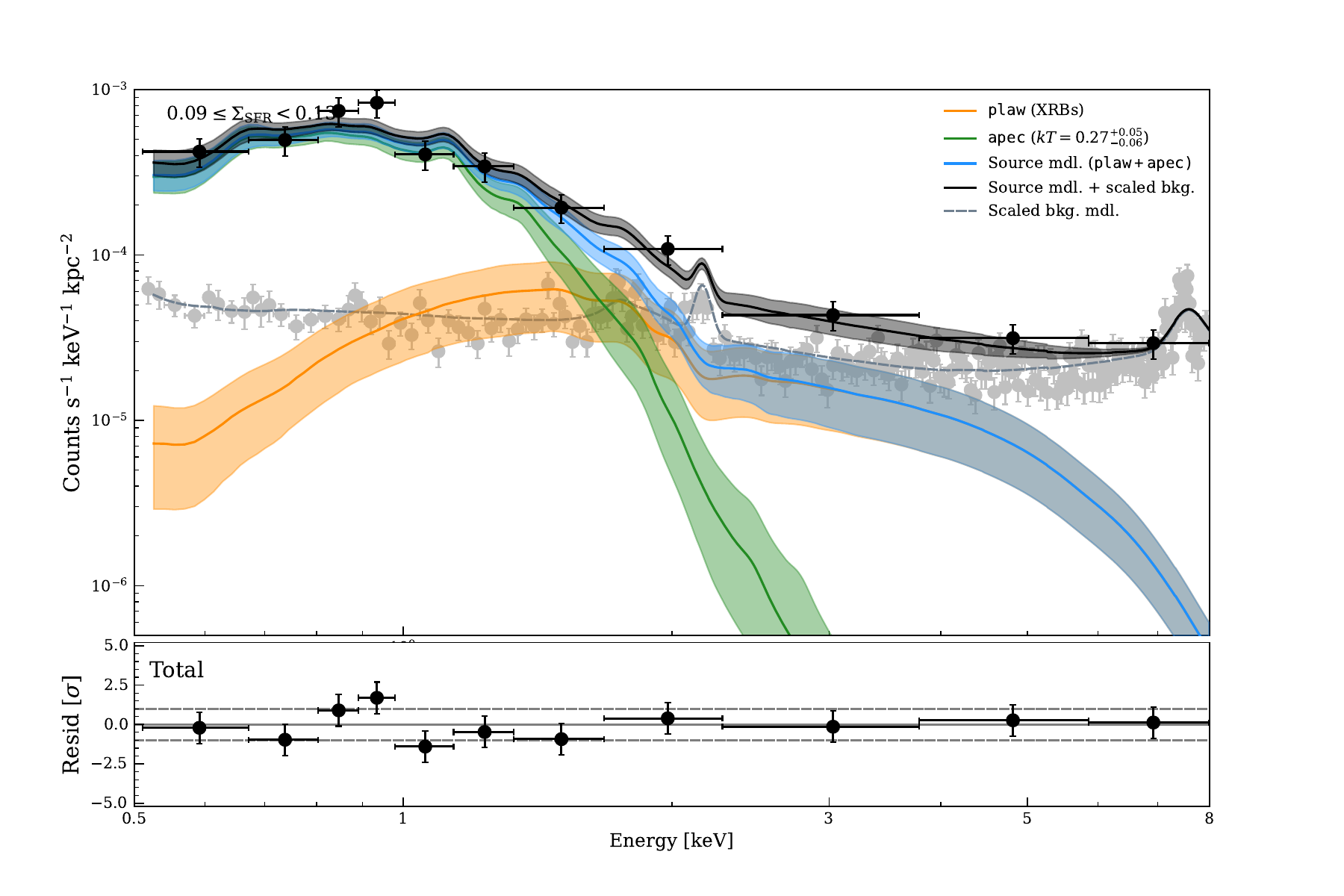}
    \includegraphics[width=0.85\textwidth]{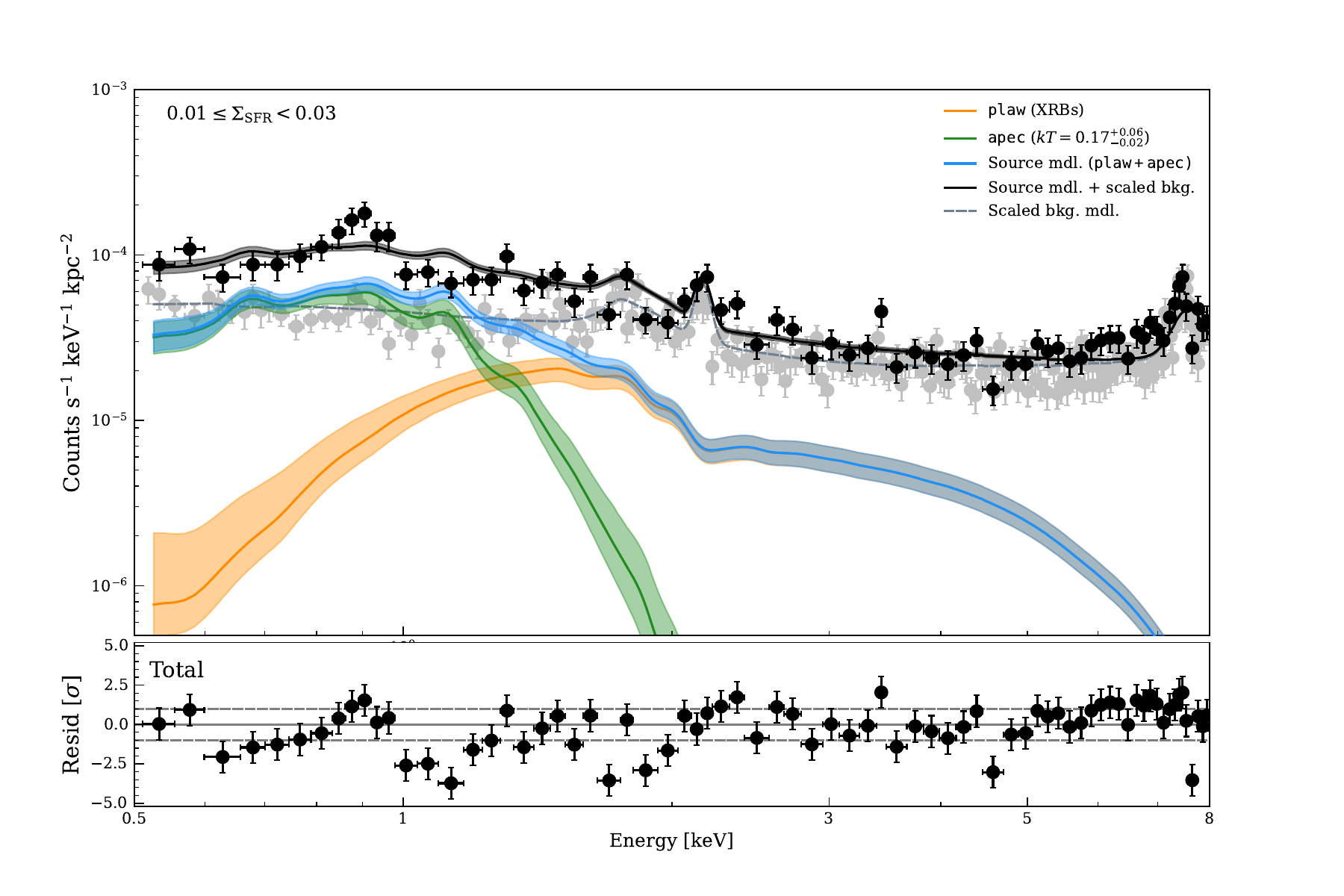}
    \caption{\label{fig:xrayfits}Bayesian X-ray spectral fits for two \SFRD\ bins: the top panel shows the fit to the two hexagonal tiles in the $0.09 \leq \SFRD / ({\rm M_{\odot}~yr^{-1}}) < 0.13$ bin, while the lower panels show the fit to the 31 hexagonal tiles in the $0.01 \leq \SFRD / ({\rm M_{\odot}~yr^{-1}}) < 0.03$ bin.  The colored bands show the 16th-84th percentile range in each model component, with the total model shown as a blank band. The central lines in each band show the median. The best-fitting background model is shown as a dashed gray line. To compactly display the fits, which were performed jointly to all five ObsIDs, we have combined the spectra and responses of all five observations and folded the fitted spectral models through the combined responses. The source spectra are displayed grouped with 10 counts per spectral bin; the background spectra are grouped with 300 counts per spectral bin.
    Spectra and models have been normalized by the total area over which the spectra were extracted to emphasize differences in intensity between different \SFRD\ bins. We show the normalized data$-$model $/ \sigma$ residuals for the best fitting model in each case, though we caution that this is a contrived method of showing the fit quality when the data are grouped for plotting, as we performed the fits to the ungrouped data with \texttt{cstat}. The remaining spectral fits are shown in \autoref{sec:specfits}.}
\end{figure*}

We configured \texttt{UltraNest} to explore the parameter space with 150 live points until 90\% of the evidence is integrated. To properly account for variations in the spectral response of Chandra over the 15 years covered by our data, we fit the ObsIDs jointly, forward-modeling the same spectral model through their different responses and combining the likelihoods, rather than combining them to derive and fit a single spectrum. The best-fitting background model for each ObsID was held constant while we fit the source model. To compactly display our fitting results, we have combined the spectra and responses using the \texttt{CIAO} \verb|combine_spectra| tool. We show the combined spectra and the spectral model in \autoref{fig:xrayfits} for the $0.09-0.13$ and $0.01-0.03~\rm M_{\odot}~yr^{-1}~kpc^{-2}$ stacks, with fits to the remaining stacks shown in \autoref{sec:specfits}.  

The thermal diffuse X-ray emission in galaxies can be modeled with two or more temperature components (modeling some more complex, unknown distribution of temperatures) on both galaxy-integrated and sub-galactic scales \citep[e.g.][]{yukita2010, yukita2012, kuntz2010, townsley2024}. In comparison to multi-temperature models, single-temperature models can also lead to the inference of larger normalizations (and consequently plasma densities) and obscuring column densities \citep{yukita2010}. We tested for the presence of a hotter component in our spectra by fitting two-temperature plasma models. We added another \texttt{apec} component to the model, and set uniform priors from $0.1-0.4$ keV and $0.4-1.0$ keV on the temperature components. We chose these priors so that the lower-temperature component encompassed the solutions from our single-temperature fits, and the two components were prevented from degenerating to a single temperature. We adopted the same prior on $\log N_H$ as the single temperature fits. We found Bayes factors marginally in favor ($\approx 2$) of two-temperature models for all bins; Bayes factors this low (i.e. ``barely worth mentioning'' on the Jeffreys scale), however, can be dominated by noise and do not necessarily indicate conclusive evidence in favor of the two-temperature model. We found that the lower-temperature component recovers the same temperature as the single temperature fit, and the normalization on the hotter component is essentially unconstrained, such that the addition of a second component does not add to our physical understanding of the plasma. We note that this finding is likely driven by the limited signal-to-noise in our spectra; spectra with more counts could reveal spectral features that conclusively require the addition of a second temperature component to fit well. For the above reasons, and also for ease of compatibility with the single-temperature \citetalias{chevalier1985} model (see \autoref{sec:wind}), we use single-temperature measurements in what follows.

\begin{figure}
    \centering
    \includegraphics[width=0.48\textwidth]{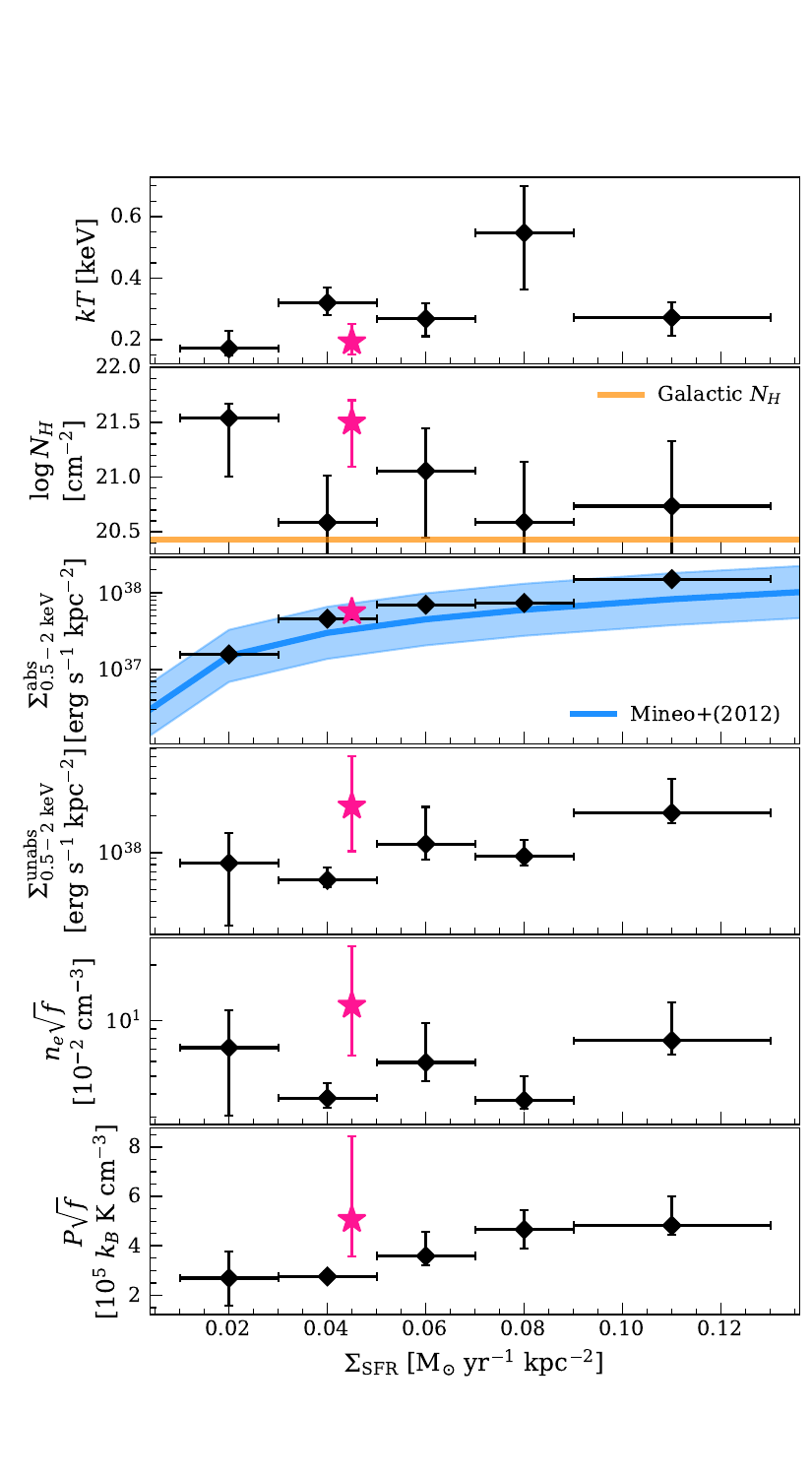}
    \caption{\label{fig:stackparams}X-ray properties from spectral fits to tiles in each of the six $\Sigma_{\rm SFR}$ bins, displayed as a function of the central $\Sigma_{\rm SFR}$ of the bin. In the second panel, the orange line shows the Galactic column density along the line of sight.
    The X-ray luminosities have been normalized as surface densities $\Sigma = L_X / A$ for fair comparison of the luminosities of the different stacks and the full galaxy. The properties measured within the X-ray extent of the galaxy are shown with a pink star marker at the average \SFRD\ of the galaxy. In the third panel we show the \citet{mineo2012} relation for the 0.5-2 keV hot gas luminosity (without absorption correction; their Equation 2) as a blue line with shaded band representing their measured 0.34 dex scatter. We have adjusted the \citet{mineo2012} relation upward (i.e., decreasing the SFR) by a factor of 1.45, accounting for the difference in IMF \citep[see section 6.5 in][]{eldridge2018}.
    }
\end{figure}

We show the median temperature, column density, absorbed 0.5$-$2 keV luminosity, and intrinsic 0.5$-$2 keV luminosity from the chains of posterior samples produced by the \texttt{BXA} procedure in \autoref{tab:stackparams} and graphically in \autoref{fig:stackparams}, in addition to the electron density, \added{volume emission measure}, and pressure of the X-ray emitting plasma. The luminosity is related to the electron density by
\begin{equation}\label{eq:ne}
    n_e\sqrt{f} = \sqrt{\frac{L_X \cos i}{\Lambda V}}
\end{equation}
where $V$ is the volume, $f$ is the filling factor of the X-ray emitting plasma, $L_X$ is the intrinsic 0.3$-$10 keV luminosity of the plasma model, and $\Lambda$ is the corresponding CIE cooling function, calculated using \texttt{pyAtomDB} v0.11.8\footnote{\url{https://github.com/AtomDB/pyatomdb/tree/master}} using the same metallicity as we adopted for the spectral fits. The electron density and pressure are sensitive to the assumption of a volume, and thus an underlying geometry, for the hot, X-ray emitting gas. We assumed a disk geometry, with a thickness of 200 pc \citep[e.g.][]{yukita2010} to calculate the volume corresponding to a hexagonal tile. Such a geometry supposes that the hot gas is confined to the plane of the galaxy, an assumption consistent with the prior we place on $N_H$. We thus have $V = 0.4~{\rm kpc^3}$ for each 1.5 kpc wide hexagon, where the X-ray emitting gas is assumed to fill a fraction $f$ of said volume. We estimated the pressure of the X-ray plasma with the ideal gas law:
\begin{equation}\label{eq:PV}
    P\sqrt{f} = n \sqrt{f} kT = 1.9 n_{\rm e} \sqrt{f} kT = 1.9 \sqrt{\frac{L_X \cos i}{\Lambda V}} k T.
\end{equation}
where $n = (n_{\rm e} + n_{\rm H} + n_{\rm He})$, and the factor of 1.9 follows from assuming all Helium in the X-ray emitting plasma is doubly ionized, with a Helium mass fraction of 30\% \added{\citep[e.g.][]{lopez2011}}.

\section{Discussion} \label{sec:discussion}

\subsection{Sub-galactic Variation of X-ray Emission} \label{sec:spatial}

We first investigate the spatial variation of the X-ray emission in the galaxy. 

\begin{figure}
    \centering
    \includegraphics[width=0.5\textwidth]{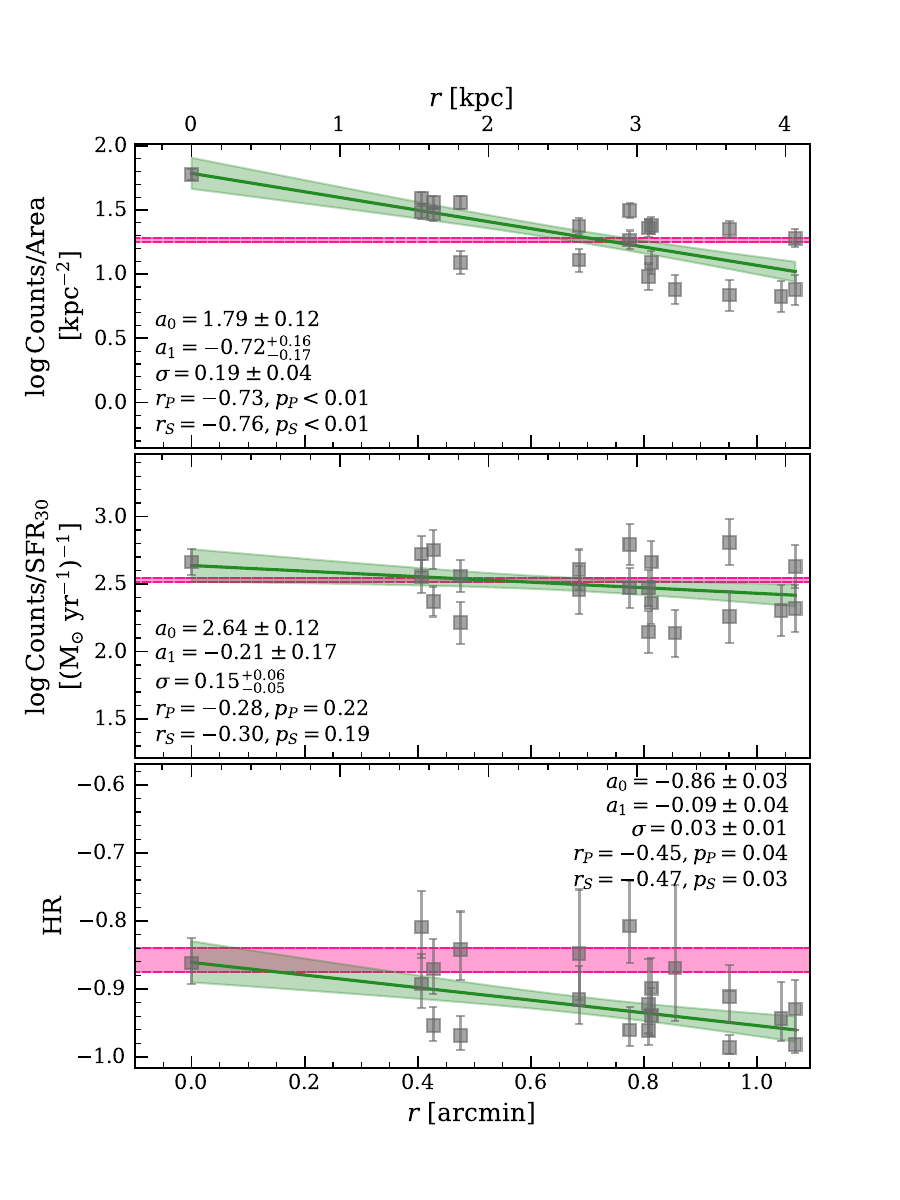}
    \caption{\label{fig:model_indep_r}Radial variation of X-ray model-independent quantities extracted in the unbinned hexagonal tiling for tiles within the measured extent of the X-ray emitting plasma (see \autoref{sec:galint}): net counts per area, net counts per $\rm SFR_{30}$ (as a proxy for X-ray production efficiency), and soft-band $(S2 - S1) / (S2 + S1)$ hardness ratio (as a proxy for temperature and obscuration; see text for definition). The radial coordinate $r$ is the distance from the center of the galaxy to the centroid of the hexagonal tile.
    The annotation for each panel gives the median and 16th-84th percentile of the fitted linear parameters ($y = a_0 + a_1 x$) with the corresponding Pearson (P) and Spearman (S) statistics and $p-$values for the correlation between the shown parameters and $r$ within the measured extent of the X-ray emitting plasma. The green line and shaded region show the pointwise median and 16th-84th percentile range of the fitted line. 
    We also calculated the same quantities in an aperture defined by the X-ray extent of the galaxy, and we show their 1$\sigma$ equivalent confidence intervals as a horizontal pink-shaded band.
    }
\end{figure}

In \autoref{fig:model_indep_r} we show the radial variation of model-independent quantities calculated in the un-binned hexagonal tiles within $2.5$ arcmin (9.5 kpc) of the center of the galaxy (i.e., within the optical extent). 
We exclude here the hexagon containing a piled-up ultra-luminous X-ray source (ULX) candidate at (R.A., Dec.) $=$ (12:18:56.10, +14:24:19.5) \citep{walton2022}. Even after point source masking with \acisx, the X-ray spectrum of the hexagon containing the candidate ULX is extremely hard, and consistent with a power law alone.

As a model-independent proxy for the hot gas X-ray production efficiency, we calculated the 0.5$-$2 keV band net counts normalized by $\rm SFR_{30}$. As our proxy for the temperature, we calculated the soft-band hardness ratio. Hardness ratio probability distribution functions (PDFs) were computed for each ObsID separately using the \texttt{FastHR} code \citep{zou2023} and combined to derive the joint PDF on the hardness ratio ${\rm HR} = (S2 - S1) / (S2 + S1)$, where $S1$ and $S2$ are the number of counts in the $0.5-1.2$ and $1.2-2.0$ keV bands, respectively. This hardness ratio should be sensitive to the temperature of the plasma (as well as the obscuration), with larger (closer to zero) values corresponding to a harder spectrum and indicating either a hotter plasma temperature or a larger obscuring column density. We see an obvious power-law trend between galactocentric distance and the surface density of counts, tracing the X-ray surface brightness profile of the disk, with the surface brightness dropping by roughly a factor of five (0.72 dex) over an arcminute. The counts per unit star formation, on the other hand, exhibit no strong trend with galactocentric distance. This proxy for the X-ray production efficiency instead varies around the galaxy-average value (see \autoref{sec:galint}) with a scatter of 0.30 dex. We see a trend, significant at the $95\%$ confidence level, between the soft-band $(S2 - S1) / (S2 + S1)$ hardness ratio and galactocentric distance, indicating harder spectra (and thus potentially higher temperatures) in the inner regions of the galaxy. This trend, however, appears to be driven primarily by the central tile (capturing the central 1.5 kpc of the galaxy) and thus could be sensitive to individually non-detected XRBs or supernova remnants. One of the two hexagonal tiles making up the 0.07-0.09 $\rm M_{\odot}~yr^{-1}~kpc^{-2}$ bin has a hardness ratio $-0.81$, harder than both the galaxy-averaged value and the value in the central hexagon, indicating possible XRB contamination or an extra source of heating (or equivalently, inefficient cooling) for the diffuse X-ray emitting gas. We discuss this particular \SFRD\ bin further below, in the context of the spectral fitting results.

\begin{figure*}
    \centering
    \includegraphics[width=0.85\linewidth]{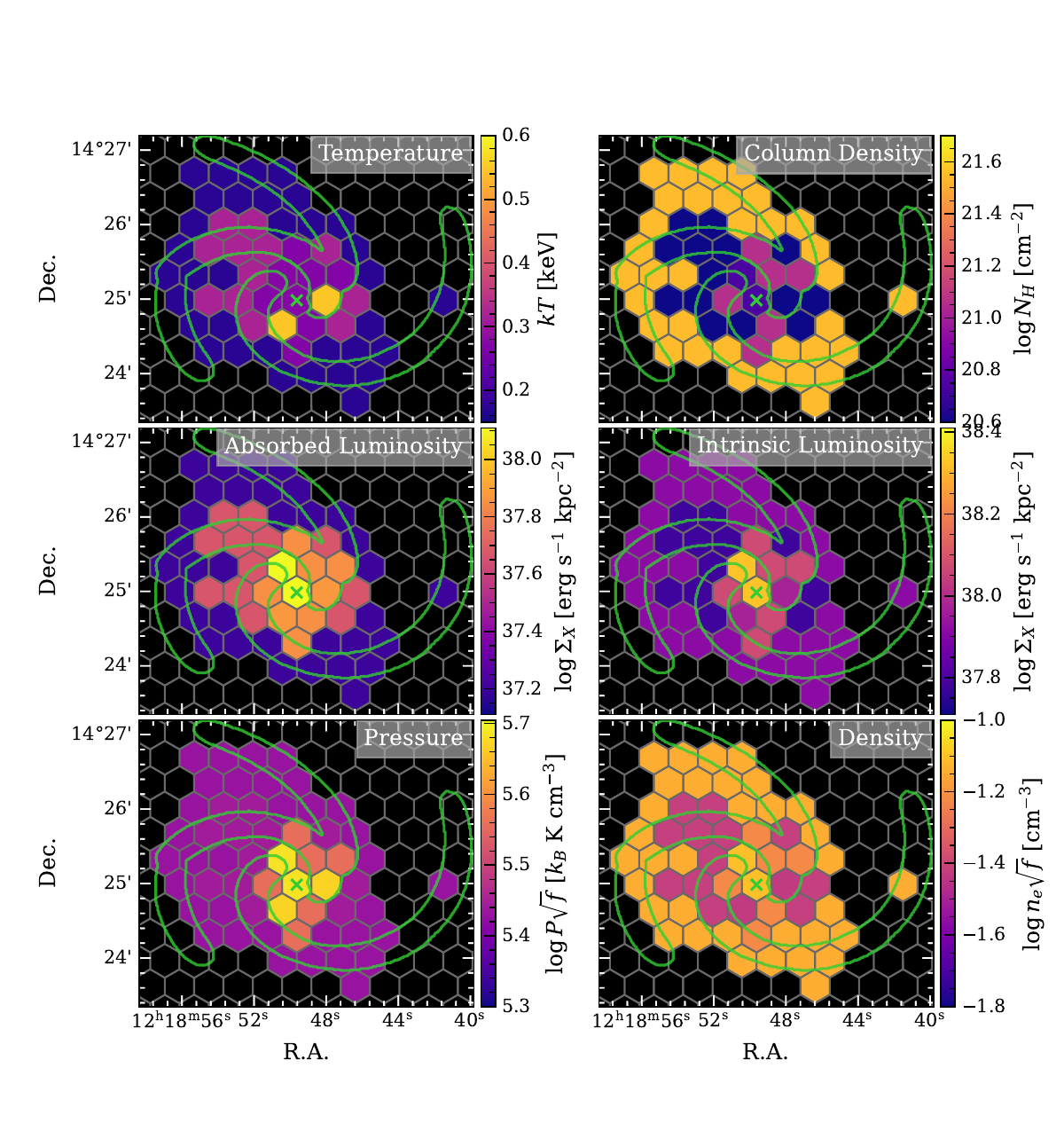}
    \caption{\label{fig:hexplot}Quantities derived from our stacked X-ray spectral fits, shown as hexagonal tile maps of the galaxy. Each tile is colored according to the value for its corresponding stack. Luminosities (second row) have been normalized as surface densities $\Sigma = L_X / A$; we show the $0.5-2$ keV luminosity of the \texttt{apec} component with and without absorption corrections. The density and pressure are calculated following \autoref{eq:ne} and \autoref{eq:PV}. 
    As a visual reference for the galaxy structure, green contours trace the locations of the spiral arms in the \citet{querejeta2021} environment masks.
    Recall that the edge-to-edge width of each hexagon is 1.5 kpc.}
\end{figure*}

The trends in the model-independent quantities are reflected in the quantities we derive from spectral fitting. We show the X-ray spectral fitting results projected onto the hexagonal tiling of the galaxy in \autoref{fig:hexplot}. This representation reveals radial variations in the plasma temperature, with hotter temperatures in the central regions. The case of the $0.07-0.09~\rm M_{\odot}~yr^{-1}~kpc^{-2}$ \SFRD\ bin is interesting: we recover a significantly hotter temperature than the other bins, finding $kT=0.55^{+0.15}_{-0.18}$ keV \added{$\sim 1.5\sigma$ higher than the mean temperature of the other bins. If, for the sake of argument, we treat the posterior $kT$ samples for all the other \SFRD\ bins as being drawn from the same underlying temperature distribution, we find a less than $0.1\%$ probability that posterior $kT$ samples from the $0.07-0.09~\rm M_{\odot}~yr^{-1}~kpc^{-2}$ bin is drawn from the same distribution (using the two-sample Anderson-Darling test), and a less than $0.1\%$ probability that the underlying temperature distributions have the same mean (using Welch's two-sample T-test). Given that the offset is statistically significant,} we checked for contamination of the diffuse emission spectrum by X-ray non-detected supernova remnants (SNR), which typically present hot, steep X-ray spectra. We searched for SNRs in the PHANGS galaxy SNR catalog \citep{li2024} that fall within this \SFRD\ bin, and masked out $1''$ radius apertures around each SNR before re-extracting the spectrum. We fit this spectrum again with a single temperature plasma model, following the same procedures given in \autoref{sec:spec}, finding that the high temperature, $kT=0.61_{-0.21}^{+0.14}$ keV, remains even after masking out counts that may be associated with SNRs. We also see that the power law XRB component is not particularly well constrained in this \SFRD\ bin, indicating that contamination from non-detected XRBs could also be to blame. The metallicity-dependent $L_X/{\rm SFR}$ relation from \citet{lehmer2024} suggests that we should expect XRBs to contribute $\log L_X = 39.2^{+0.3}_{-0.4}~\rm erg~s^{-1}$ in this bin. The point sources that we subtracted with \acisx\ account for $2\times10^{38}~\rm erg~s^{-1}$, and the power law component of the spectral fit adds a further $4.5^{+5.5}_{-4.2}\times 10^{37}~\rm erg~s^{-1}$, for a total of $\log L^{\rm XRB}_X = 38.5^{+0.1}_{-0.2}$. \added{Unresolved XRBs thus appear the most likely explanation, given the shortfall between the total XRB luminosity we can account for and the model prediction. We note, however that} this is a sub-galactic region, which necessarily has an incompletely sampled XLF, such that the observed XRB luminosity might be significantly lower than the luminosity inferred from the scaling relation \added{\citep[e.g.][]{lehmer2024}}. \added{Though we consider them the most likely culprit,} it is thus difficult to conclusively attribute the higher temperature to XRBs. We may therefore be observing the true emission from the hot gas, shock-heated to higher temperatures than neighboring regions of the galaxy \added{(potentially due to recent interactions with other cluster member galaxies) or inefficiently cooling due to lower metal abundances. In the absence of more photons we are unable to conclusively determine a cause for this anomalously high temperature. Regardless of the cause, the large uncertainty on the temperature means that this bin carries relatively little statistical weight in what follows.}

We see that the column density peaks $4-6$ kpc from the center, reflecting the peaks in the 21 cm emission map (see \autoref{sec:21cm_image}, \autoref{fig:21cm_image}), which have been averaged over our large spectral extraction regions. This is consistent with our prior on $N_H$, though we note again that the 21 cm-based prior is quite broad. The density, directly derived from the absorption corrected luminosity ($\propto L^{1/2}$), naturally follows the same profile, while the pressure ($\propto L^{1/2} T$) shows similar spatial variations as the plasma temperature. We note that we assumed a constant disk thickness and filling factor to calculate the density and pressure of the hot gas, \added{such that a factor of two increase (decrease) in the disk thickness decreases (increases) the density and pressure throughout the disk by a factor of $\sqrt{2}$.} If instead we assumed an exponentially declining profile for the hot gas, we would observe larger densities in the outskirts of the galaxy due to the declining volume, and consequently a flattening of the pressure profile seen in \autoref{fig:stackparams}. A flared disk with a larger height in the outskirts of the galaxy would produce the opposite effect: a larger volume in the outskirts would reduce the pressure. However, flaring is typically observed in the outermost disk, outside of where we are able to produce X-ray constraints. Due to our lack of information on the true geometry of the X-ray emitting plasma, we continue to treat it as a constant thickness disk.

Given that the \SFRD\ decreases rapidly with galactocentric distance, the sub-galactic relationships between the hot gas properties and \SFRD\ shown in \autoref{fig:stackparams} can also be understood as 1D projections of the radial trends described above. In the third panel of \autoref{fig:stackparams}, we compare our measurements to the commonly-used \citet{mineo2012} hot gas $L_X-{\rm SFR}$ relation, finding that our sub-galactic measurements are remarkably consistent with the relation, and follow roughly the same trend. This, along with the nearly-constant counts/SFR$_{30}$ ratio seen in \autoref{fig:model_indep_r}, suggests that the processes which create the tight $L_X-{\rm SFR}$ relation and maintain a nearly-constant X-ray production efficiency may operate on sub-galactic scales; we explore this further in \autoref{sec:wind}.

\subsection{Galaxy-Integrated Properties}\label{sec:galint}
To provide a galaxy-integrated comparison for the sub-galactic X-ray properties, we measured the extent of the X-ray emitting gas, following the procedures from \citet{smith2019}. We retrieved the parameters of the $B-$band 25 mag arcsec$^{-2}$ isophote (the ``D25 ellipse''; $a=150\farcs3$, $b=140\farcs1$) from the HyperLeda database\footnote{\url{http://atlas.obs-hp.fr/hyperleda/search.html}}, and measured the X-ray radial profile of the galaxy from the merged, point-source-subtracted, exposure-corrected 0.5$-$2 keV image created with \verb|merge_obs| in \autoref{sec:data:xray}. We use 15 linearly spaced elliptical annuli sized between $10''$ and the D25 ellipse to construct the radial profile, subtracting an estimate of the background flux we obtained from an elliptical annulus with inner radii equal to the D25 isophote and outer radii scaled up by 20\%. Following \citet{smith2019}, we take the extent of the X-ray emission as the point where the surface brightness reaches $3\times10^{-9}$ photons s$^{-1}$ cm$^{-2}$ arcsec$^{-2}$, \added{which \citet{smith2019} found produced apertures enclosing roughly 90\% of the total flux from the galaxy in a heterogeneous sample.} Based on this criterion\footnote{Notably \citet{smith2019} used a 0.3$-$1 keV map to estimate the extent, while we use a 0.5$-$2 keV map; we find that their surface brightness criterion still gives a reasonable estimate in our case.}, the extent along the major axis is \angextent, corresponding to \physextent\ kpc. Within this X-ray extent, the total star formation rate is ${\rm SFR}_{\rm 30} = 3.0\pm0.2~{\rm M_{\odot}~yr^{-1}}$, and $\log M_{\star}/{\rm M_{\odot}} = 10.04 \pm 0.02$. The X-ray extent contains 973 net counts, corresponding to 18.3 counts kpc$^{-2}$. We calculated the hardness ratio following the same procedure as we used for the hexagonal tiles, finding a galaxy-averaged $\rm HR = -0.86\pm0.02$, and we show them alongside the sub-galactic measurements in \autoref{fig:model_indep_r}.

We extracted spectra from the point-source subtracted event lists inside the region defined above. 
We fit a single temperature model to the X-ray spectrum, finding $kT = $ \kTextent\ keV and an absorbed 0.5$-$2 keV luminosity \LXSBabsextent\ erg s$^{-1}$; \added{the fit is shown in \autoref{fig:specfits_extra_3}}. The luminosity corresponds to a $L_X/{\rm SFR_{30}}$ ratio $1.2\times10^{39}~{\rm erg~s^{-1} (M_{\odot}~yr^{-1})}$, consistent with the adjusted \citet{mineo2012} relation plotted in the third panel of \autoref{fig:stackparams}. \added{The single temperature emission measure is $n_e^2 V = 40.6^{+136}_{-29.1} \times 10^{-3}~\rm cm^{-6}~kpc^{3}$, similar to the range of emission measures \citet{mineo2012} reported for galaxies with similar SFR (see their Table 8). The temperature we recover is also consistent with the 0.24 keV average temperature of the \citet{mineo2012} sample of galaxies.} Given the larger number of counts available, we adopted an uninformative prior on $\log N_H$ (rather than the 21 cm-based priors we assumed for sub-galactic fitting), finding $\log N_H /{\rm cm^{-2}} = 21.5^{+0.2}_{-0.4}$, corresponding to an absorption-corrected 0.5$-$2 keV luminosity \LXSBunabsextent\ erg s$^{-1}$. This column density is in line with our fits to the two lowest-\SFRD\ bins (which cover the largest physical area, and contain the 21 cm emission peaks)\added{, and we find that our measured column density is consistent with the galaxy-integrated column densities of the \citet{mineo2012} sample, which were found to typically be on the order of a few times $10^{21}$ cm$^{-2}$ (see their Table 3).} We again tested the inclusion of an additional, hotter $0.4-1$ keV component, finding that its normalization is a factor of $\sim1000$ lower than the lower temperature component, such that we recover an emission-measure-weighted temperature $\langle kT\rangle =$ \kTextenttwo\ keV. Given that such a hotter component appears to be weak, we used the single-temperature fit to derive a density and temperature for the entire galaxy, following \autoref{eq:ne} and \autoref{eq:PV}. We find $n_e\sqrt{f} = 11.0^{+11.9}_{-5.1}\times10^{-2}{\rm cm^{-3}}$ and $P\sqrt{f} = 4.60^{+3.05}_{-1.36}\times10^{5}~{\rm k_B~K~cm^{-3}}$, noting that assuming a filling factor equal to 1 corresponds to the volume-averaged pressure. \added{This density is on the high end compared to \citet{mineo2012}, who estimated $10^{-3}-10^{-2}~\rm cm^{-3}$ for their sample under the assumption of a larger volume occupied by the X-ray emitting plasma, based on Chandra observations of edge-on galaxies (including outflowing galaxies) by \citet{li2013} -- see our brief discussion of the plasma volume and filling factor in \autoref{sec:spatial}.}

\subsection{Sub-galactic Wind Model}\label{sec:wind}

To provide a physical interpretation to our constraints on the sub-galactic X-ray properties, we interpret the results of our sub-galactic UV-IR SED fitting and X-ray spectral fitting in the context of the \citetalias{chevalier1985} model. We derive expressions for the temperature and X-ray luminosity of an adiabatically expanding hot plasma using the \citetalias{chevalier1985} model in \autoref{sec:cc85}.
The model predictions for $kT$ and $L_X$ depend on three unknown parameters: $\epsilon$, the thermalization efficiency (see \autoref{eq:Edot0}), $\beta$, the mass loading factor (see \autoref{eq:Mdot}), and $R_{100}$, the size of the idealized spherical starburst into which energy and mass are deposited by CCSNe. An increase in $\epsilon$ corresponds to an increase in the fraction of the energy from CCSNe which is thermalized in shocks, and an increase in $\beta$ corresponds to an increase in the amount of ISM mass entrained by the ejecta into the wind fluid.

To quantitatively examine the variation of $kT$ and $L_X / {\rm SFR_{30}}$ and to constrain the wind parameters in the galaxy on sub-galactic scales, we assume simple forms for $\beta$, $\epsilon$, and $R_{100}$ as functions of $\Sigma_{\rm SFR}$:
\begin{align}\label{eq:model}
    &\beta = 1 + (\beta_0 - 1) (\Sigma_{\rm SFR} / \Sigma_0)^{-\gamma_\beta} \\
    &\epsilon = \epsilon_0 \\
    &R_{100} = R_0 (\Sigma_{\rm SFR} / \Sigma_0)^{\gamma_R},
\end{align}
where the functional form for $\beta$ is such that $\beta > 1$. We focus on cases where $\epsilon$ is constant with $\Sigma_{\rm SFR}$, since $L_X / {\rm SFR}$ is more sensitive to changes in $\beta$. We investigated four possible scenarios for the variation of parameters with $\Sigma_{\rm SFR}$, described in \autoref{table:windspec}. In all cases, we allow the level of $\epsilon$ to vary and modify the scaling of $\beta$ and $R$ with \SFRD. We first assume that $\beta$ is constant, then that $\beta \propto \SFRD^{-1/3}$ (the scaling imposed by the upper limit on $\beta$; see \autoref{eq:betamax}), and then allow $\beta$ to scale with an arbitrary exponent. As a final test, we allow both $\beta$ and $R$ to scale with arbitrary exponents. In all cases with non-constant $\beta$, it is defined as a decreasing function of $\Sigma_{\rm SFR}$ (i.e. $\gamma_\beta > 0$), which is supported both by the upper limit on $\beta$ derived from the assumption of adiabatic expansion and by observations \citep[e.g.][]{zhang2014}.

To calculate model $L_X/{\rm SFR}$ and $\beta_{\rm max}$ according to \autoref{eq:LXSFR} and \autoref{eq:betamax}, we calculated cooling functions using \texttt{pyAtomDB}. To account for radiative cooling outside of the Chandra bandpass, we use a ``bolometric'' cooling function calculated over the 0.01$-$100 keV energy range to compute $\beta_{\rm max}$.

\begin{deluxetable*}{lcccchcccc}
    \tablecaption{\label{table:windspec}Priors assumed for each of the cases of the galactic wind model.}
    \tablehead{
    \colhead{Case} &
    \colhead{Descr.} &
    \colhead{$\beta_0$} &
    \colhead{$\gamma_\beta$} & 
    \colhead{$\epsilon_0$} & 
    \nocolhead{$\gamma_\epsilon$} &
    \colhead{$R_0$} & 
    \colhead{$\gamma_R$} &
    \colhead{$\Sigma_0$} &
    \colhead{$N_{\rm param}$}
    }
    \startdata
    1 & $\beta = \rm const$ & $\mathcal{U}(1,100)$ & 0 & $\mathcal{U}(0.01,1)$ & 0 & 1 & 0 & 1 & 2\\
    2 & $\beta \propto \SFRD^{-1/3}$ & $\mathcal{U}(1,40)$ & 0.33 & $\mathcal{U}(0.01,1)$ & 0 & 1 & 0 & 1 & 2\\
    3 & $\beta \propto \SFRD^{-\gamma_{\beta}}$ & $\mathcal{U}(1,40)$ & $\mathcal{U}(0.0,1.0)$ & $\mathcal{U}(0.01,1)$ & 0 & 1 & 0 & 1 & 3\\
    4 & $\beta \propto \SFRD^{-\gamma_{\beta}}, R \propto \SFRD^{\gamma_{R}}$ & $\mathcal{U}(1,40)$ & $\mathcal{U}(0.0,1.0)$ & $\mathcal{U}(0.01,1)$ & 0 & $\mathcal{U}(0.5,7)$ & $\mathcal{N}(0.5,0.15)$ & 1 & 5\\
    \enddata
    \tablecomments{$\mathcal{U}(a,b)$ is the Uniform distribution on the interval $[a,b)$;  $\mathcal{N}(\mu,\sigma)$ is the normal distribution with mean $\mu$ and standard deviation $\sigma$. Parameters with a single value are fixed.}
\end{deluxetable*}


We simultaneously fit $L_X / {\rm SFR}$ and $kT$ for the five sub-galactic $\Sigma_{\rm SFR}$ bins
using our model defined in Equations \ref{eq:model}--9 and the predictions from \autoref{eq:kT} and \autoref{eq:LXSFR}, adopting half of the 16th-84th percentile range from the \texttt{BXA} chains as the uncertainties on $L_X$ and $kT$. We 
incorporate the width of the $\Sigma_{\rm SFR}$ bins into the uncertainty on $L_X / {\rm SFR}$. We calculate $\nu$, the number of CCSNe per solar mass formed, from the tables provided with the BPASS release, summing the number of CCSNe over the same time interval as we calculate the SFR. For our BPASS v2.2.1 stellar population models, assuming a \citet{chabrier2003} IMF, $\nu = 8.2\times10^{-3}~{\rm M_{\odot}^{-1}}$ over the $0-30$ Myr stellar age interval.

\begin{figure*}
    \centering
    \includegraphics[width=0.85\textwidth]{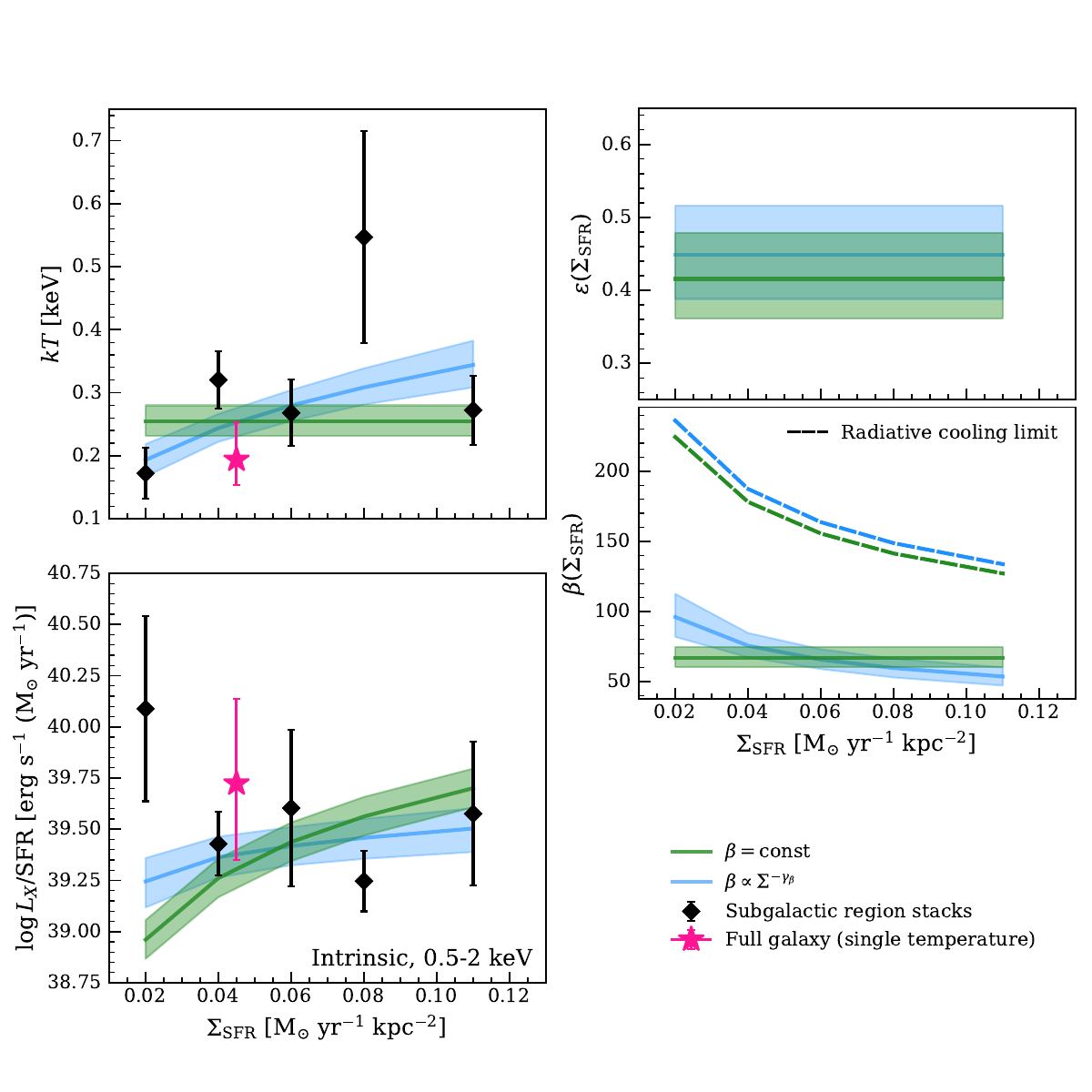}
    \caption{\label{fig:subscaling} \added{We show the results of fitting the simple sub-galactic wind models described in \autoref{sec:wind} to the X-ray properties of the \SFRD\ bins.}
    The left column shows $kT$ and $\log Y = \log L_X / {\rm SFR}$ for each \SFRD\ bin.
    Each point shows the median and 16th-84th percentile range, where the bin width has been added appropriately to the uncertainty on $\log L_X/{\rm SFR}$. The effective temperature and $\log L_X/{\rm SFR}$ for the whole galaxy from a single temperature fit to the spectrum extracted from the region defined in \autoref{sec:spatial} is shown as a magenta star. In each panel, we show posteriors on the galactic wind model fit to the sub-galactic regions as lines and shaded regions representing the posterior median and the 16th$-$84th percentile ranges, respectively, where the panels in the right column show the resulting $\epsilon$ and $\beta$ as a function of \SFRD. The second panel in the right column also shows the upper limit on $\beta$ given by \autoref{eq:betamax} as a dashed line for each fit. We show Case 2 with $\beta = {\rm const}$ in green and Case 3 with $\beta \propto \SFRD^{-\gamma_\beta}$ in blue. Case 1 is omitted due to its similarity to Case 2, and Case 4 is omitted due to poor constraints.}
\end{figure*}

We adopted priors on the free parameters for each case as shown in \autoref{table:windspec}. We sampled the posterior distributions with \texttt{emcee}, using an ensemble of 64 walkers and running them until we obtained 1000 independent samples from the posterior. The fitted parameters are given in \autoref{tab:fitparams}, and \autoref{fig:subscaling} shows the posteriors on the models for three cases. We compare the reduced $\chi^2$, Akaike information criterion (AIC), and Bayesian information criterion (BIC) in \autoref{tab:fitparams}, where the latter two are measures of the goodness of fit penalized by the number of estimated parameters (i.e., model complexity). Heuristically, the model with lower AIC or BIC should be preferred. In our case, this indicates Case 2, where $\beta \propto \Sigma^{-1/3}$, best explains the observations with the least complicated model. We also compared models pairwise with the Bayesian posterior odds ratio; with this method we find a \added{very} weak preference for Case 3 \added{($\ln P_2 /P_3 = -0.02$)}, where $\beta \propto \Sigma^{-\gamma_\beta}$ with $\gamma_\beta$ a free parameter. \added{The information criteria slightly favor Case 2, with $\Delta\rm AIC = -1.97$ and $\Delta\rm BIC = -2.27$ compared to Case 3.} However, the two cases yield consistent results: in Case 3, we find  $\gamma_\beta = 0.34_{-0.09}^{+0.11}$. In what follows, we adopt Case 3 as the preferred model \added{over the $\gamma_\beta = 0.33$ special case favored by AIC and BIC, so that we can examine the range of power law slopes consistent with the data and the resulting impact on the variation of X-ray production efficiency with \SFRD.} In all cases, we recover thermalization efficiencies in the neighborhood of 50\%: slightly less than half of the kinetic energy from CCSNe is converted to internal thermal motions of the shocked gas. Since mass loading acts to slow down the galactic winds, the drop in mass loading with increasing $\Sigma_{\rm SFR}$ implies a corresponding increase in the speed of the wind as star formation intensifies, where the supernovae ejecta encounter and entrain less interstellar gas. This is somewhat counter-intuitive, as larger star formation rate densities are associated with increased density of molecular gas \citep[e.g.][]{kennicutt1998}. The drop in mass loading may thus correspond, physically, to the expansion of supernova ejecta into volumes already cleared by other supernovae, given the larger supernova rates ($\propto {\rm SFR}$) in highly star-forming regions. \added{The TIGRESS simulations \citep{ko2017} of spiral galaxies include a similar effect; in Figure 3 of \citet{kko2020}, for example, we see large volumes of hot, low density gas created by clustered, overlapping supernovae.} We note also that even when $\beta \propto \SFRD^{\approx -1/3}$, the bulk mass flow into the ISM increases with SFR (and \SFRD), since $\dot M = \beta{\rm SFR}$.

\begin{deluxetable*}{cccccccccc}
\tablecaption{\label{tab:fitparams}Model comparison statistics and sampled quantities for each wind model scenario.}
\tablehead{
\colhead{Case} & 
\colhead{Descr.} &
\colhead{AIC ($\Delta$AIC)\tablenotemark{a}} & 
\colhead{BIC ($\Delta$BIC)\tablenotemark{a}} & 
\colhead{$\chi_{\nu}^2$} & 
\colhead{$\beta_0$} & 
\colhead{$\gamma_\beta$} & 
\colhead{$\epsilon_0$} & 
\colhead{$R_0$} & 
\colhead{$\gamma_R$}}
\startdata
1 & $\beta = \rm const$ & 25.78 (8.96) & 26.38 (8.96) & 2.72 & $67.6_{-7.2}^{+7.8}$ & $0.00$ & $0.42_{-0.06}^{+0.07}$ & $1.0$ & $0.00$ \\
2 & $\beta \propto \SFRD^{-1/3}$ & 16.82 (0.00) & 17.42 (0.00) & 1.60 & $26.7_{-2.7}^{+3.1}$ & $0.33$ & $0.45_{-0.06}^{+0.07}$ & $1.0$ & $0.00$ \\
3 & $\beta \propto \SFRD^{-\gamma_{\beta}}$ & 18.79 (1.97) & 19.69 (2.27) & 1.83 & $26.1_{-7.0}^{+8.0}$ & $0.34_{-0.09}^{+0.11}$ & $0.45_{-0.06}^{+0.07}$ & $1.0$ & $0.00$ \\
4 & $\beta \propto \SFRD^{-\gamma_{\beta}}, R \propto \SFRD^{\gamma_{R}}$ & 19.56 (2.74) & 21.07 (3.65) & 1.91 & $27.8_{-7.9}^{+8.3}$ & $0.31_{-0.10}^{+0.10}$ & $0.44_{-0.12}^{+0.14}$ & $4.5_{-2.0}^{+1.8}$ & $0.51_{-0.10}^{+0.10}$ 
\enddata
\tablenotetext{a}{In the AIC and BIC columns, the difference in the statistic from the minimum is shown in parentheses.}
\end{deluxetable*}

The $L_X/{\rm SFR}$ ratio is commonly recast in the literature to a dimensionless X-ray production efficiency $\eta = L_X / \dot E_0$, which weighs the X-ray radiative cooling rate against $\dot E_0$, the bolometric energy injection rate into the ISM (see \autoref{eq:Edot0}).  \autoref{eq:LXSFR} suggests that $\beta\propto\SFRD^{-1/3}$ will produce $\eta \sim {\rm constant}$. However, as a result of our parametrizations, $kT$ is a function of \SFRD, and the temperature-dependence of $\Lambda$ produces a residual, shallow correlation between $\eta$ and \SFRD. We find, for our adopted Case 3, 
\begin{align}
    kT &= (0.72^{+0.26}_{-0.18}) \SFRD^{0.34\pm 0.10}~[\rm keV]\\
    \eta &= (0.03^{+0.02}_{-0.01}) \SFRD^{0.34 \pm0.18}
\end{align}

\begin{figure}
    \centering
    \includegraphics[width=0.5\textwidth]{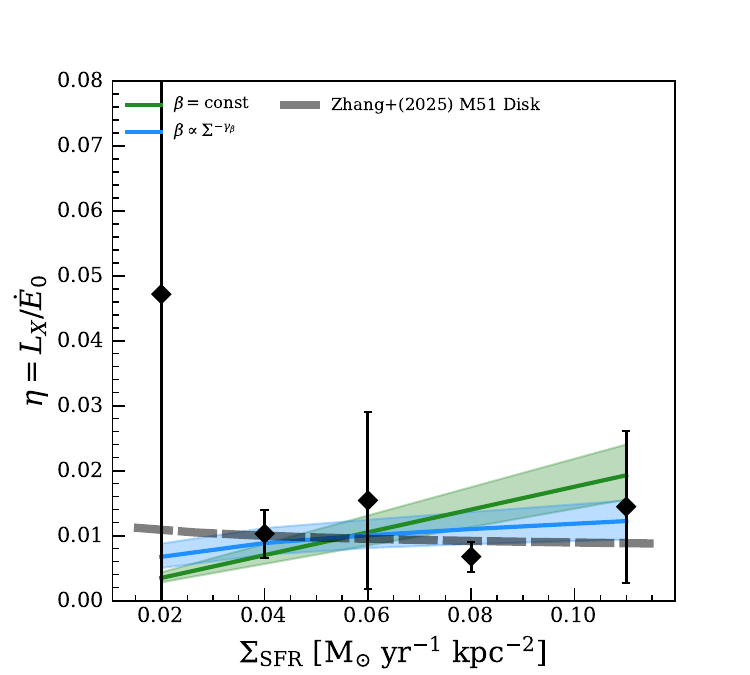}
    \caption{\label{fig:eta}The $L_X/{\rm SFR}$ ratio measured for the \SFRD\ bins (as in the second panel of \autoref{fig:subscaling}, re-cast as the X-ray radiative efficiency $\eta = L_X/\dot E_0$). We show the same model fits as in \autoref{fig:subscaling} with colored lines, alongside the $\eta$ calculated for each of the $\Sigma_{\rm SFR}$ bins as black diamonds. We also plot the power law relationship derived by \citet{zhang2025} for the disk (thick gray line) of M51. Our data agree reasonably well with the \citet{zhang2025} fit, which was derived by over a similar range of \SFRD. However, our wind model fits can only recover the negative power law slope they predict for the disk if we allow $R_{100}$ to increase as a function of $\Sigma_{\rm SFR}$; such a model could be supported or ruled out by a larger sample, enabling a finer set of $\Sigma_{\rm SFR}$ bins with smaller uncertainties.
    }
\end{figure}

\citet{zhang2025} recently measured $\eta \propto \SFRD^{0.82}$ for the center of M51, much steeper than our results. In the context of our modeling, their result is consistent with a constant mass loading factor in the center of the galaxy: when we fix $\beta$, we find $\eta = (0.18 \pm 0.04) \SFRD$. Conversely, for the outer disk of M51, \citet{zhang2025} found $\eta \propto \SFRD^{-0.12}$. Our models can only reproduce such a scaling when we allow $R$ to increase with \SFRD: our fit with $R\propto\SFRD^{\gamma_R}$ yields $\eta = 0.01^{+0.005}_{-0.003}\SFRD^{-0.10^{+0.18}_{-0.17}}$. The different $\eta$ trajectories predicted by all of our model fits are presented in \autoref{fig:eta}. Our model fitting is inconclusive about the large-scale relationship between the X-ray production efficiency and star formation: while our preferred model suggests the two are weakly coupled, with the production of X-rays from hot gas becoming slightly stronger per unit star formation in the most active regions of the galaxy, the large uncertainties on the data also allow stronger and weaker correlations, and the possibility that star formation is anti-correlated with the X-ray production efficiency over at least part of the galaxy.

\citet{zhang2025} interpreted the two different power law slopes they found in the central region and disk of M51 by suggesting that CCSNe are not the dominant contributors to the thermalization of the ISM throughout the entirety of M51: in the youngest, most massive star clusters, winds from massive stars will be the dominant source of energy (and mass) injection into the ISM \citep[e.g. the central star cluster of the Tarantula nebula;][]{townsley2024} due to the $\sim 3$ Myr time delay between the onset of star formation and the first CCSNe. In contrast, we find that the temperatures and luminosities we measure (averaged across kpc-scale regions of the galaxy) are consistent with a model including only energy and mass injection from CCSNe. However, the range of \SFRD\ we can cover in our kpc-scale analysis of this galaxy is limited. \citet{zhang2025} also derived their steep ``center'' relation on a scale comparable to a single one of our hexagonal tiles: their central region covers the inner $50''$ of M51, approximately 2 kpc, and only significantly departs from our observations for $\SFRD \gtrsim 0.1~\rm M_{\odot}~yr^{-1}~kpc^{-2}$. The fact that we do not observe such a steep correlation between $\eta$ and \SFRD\ could thus be an effect of the increased distance to NGC 4254 and limited signal-to-noise on sub-kpc scales -- combined with the lack of a clear central starburst with $\SFRD \gtrsim 0.1$ -- limiting our ability to probe the high-\SFRD, low galactocentric distance regime.

\subsection{The Hot ISM in Context} \label{sec:context}

In \autoref{fig:ISM_composite} we show a three-color composite of the three different phases of the ISM probed by these data: the cold, dense, molecular ISM with ALMA in red; the warm, dense ionized ISM with MUSE in green; and the hot, diffuse ISM with Chandra in blue. On galaxy integrated scales, we should expect that the soft X-ray emission from hot gas correlates with molecular gas mass and H$\alpha$ due to the combination of the tight relationship between X-ray luminosity and SFR with the correlations between SFR and molecular mass and SFR and H$\alpha$ luminosity. However, \added{qualitative inspection of the lower panel of} \autoref{fig:ISM_composite} reveals a more complex picture: most bright X-ray features are associated on \added{$\sim 1$ kpc scales} with star-forming complexes traced in H$\alpha$, but the inverse is not true. Likewise, \added{outside the central kpc of the galaxy the X-ray bright features are not consistently located near molecular cloud complexes traced by CO, and} the hot gas emission appears to be associated with voids in the CO emission in \added{at least one region at R.A.$=$12:18:52.2, Dec.$=+$14:25:10.0}. \added{A more quantitative assessment of sub-kpc spatial associations between diffuse X-ray and multiwavelength features is not possible here due to the limited number of X-ray photons and the limits of our astrometry (see \autoref{sec:data:xray}).} 

\added{However, even though} we are only able to measure the intrinsic properties of the X-ray emitting plasma on scales $>1.5$ kpc, we might expect these properties to be connected to the other phases of the ISM at both 1.5 kpc scales and small scales ($\approx 100$s of pc), where supernova winds are driven into and shock-heat the ISM. As a simple investigation of the multi-scale, multi-phase correlations of the ISM, we perform correlation analyses between the multi-scale molecular properties compiled in the \citet{sun2023} tables, multi-scale H$\alpha$ luminosity measurements, and the properties of the hot gas traced in X-rays.

To complement the multi-scale molecular properties, we derive analogous H$\alpha$ measurements. We first computed the attenuation-corrected H$\alpha$ luminosity over each 1.5 kpc hexagon, which we denote $L^{1.5~\rm kpc}_{\rm H\alpha}$. For each hexagon, we also searched the \citet{groves2023} nebular catalog for H II regions contained within the hexagon, and computed the weighted average of their attenuation-corrected H$\alpha$ luminosities, such that the contributions of poorly-detected nebulae are diluted. We denote this quantity $\langle L^{\rm H~II}_{\rm H\alpha}\rangle$: the average H$\alpha$ luminosity of individual H II regions. We also calculate the total H$\alpha$ luminosity of all detected H II regions in each hexagon, denoted $L^{\rm H~II}_{\rm H\alpha}$. We note that the luminosity-weighted average sizes of these H II regions and H II region complexes suggest we are probing linear scales $100-200$ pc, comparable to the cloud-scale measurements of the molecular gas mass. However, many of the H II regions are unresolved, giving us only upper limits on their sizes. As such we do not compute H$\alpha$ surface brightnesses.

The soft band net counts are significantly correlated with both the H$\alpha$ and the molecular gas mass surface density on 1.5 kpc scales, implying a correlation with the brightness of the diffuse X-ray emission (though the net counts are not corrected for absorption, and we have not attempted to remove contributions from non-detected X-ray binaries). We might expect this to be the case given that $\rm H\alpha$ and molecular gas mass density both correlate with the star formation rate, which correlates with the X-ray surface brightness. \added{The contribution of non-detected X-ray binaries may artificially strengthen the correlation, but we have seen in spectral fits that the soft counts are dominated by the plasma component in our point-source-masked data.} Interestingly, the strength and significance of the correlation between the net counts and the H$\alpha$ luminosity are dependent on the method we use to calculate the H$\alpha$ luminosity: the strongest and most significant (Spearman $r=0.84$, $p=8\times10^{-6}$) correlation is between the net counts and $L^{1.5~\rm kpc}_{\rm H\alpha}$, whereas the net counts are not well-correlated with $\langle L^{\rm H~II}_{\rm H\alpha}\rangle$ (Spearman $r=0.28$, $p=0.25$). We see that the total H II region luminosity $L^{\rm H~II}_{\rm H\alpha}$ is also correlated with the net counts, albeit less strongly and significantly than the total luminosity: Spearman $r=0.77$, $p=1\times10^{-4}$. These three correlations would seem to indicate that the brightest star-forming complexes alone are not responsible for producing the CCSNe which shock-heat the X-ray emitting ISM, and that the non-detected star-forming regions included in $L^{1.5~\rm kpc}_{\rm H\alpha}$ but not $L^{\rm H~II}_{\rm H\alpha}$ play some role in heating the ISM. We see similar multi-scale correlations with the molecular gas mass density: $\Sigma^{1.5~\rm kpc}_{\rm mol}$ is more strongly and significantly correlated (Spearman $r=0.50$, $p=0.03$) with the net X-ray counts than $\langle \Sigma^{\rm 150~pc}_{\rm mol}\rangle$ (Spearman $r=0.45$, $p=0.05$). These weak correlations reflect the complex interrelation between molecular gas and X-ray plasma visible in \autoref{fig:ISM_composite}, and motivate further study of the sub-galactic relationship between these phases of the ISM.

\begin{figure}
    \centering
    \includegraphics[width=0.50\textwidth]{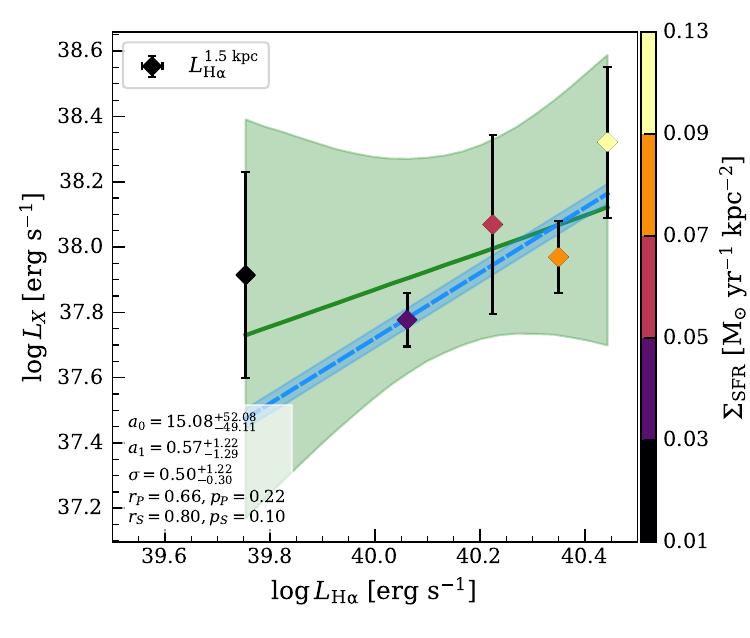}
    \caption{\label{fig:muse_scaling} MUSE-derived scaling relations for $L_X$ with the $\rm H\alpha$ luminosity, averaged over the \SFRD\ bins in which we extracted X-ray spectra. We show the 1.5 kpc-scale $\rm H\alpha$ luminosity as diamond markers, where the face color of the markers corresponds to the \SFRD\ of the bin.
    We show the pointwise median and 16th-84th percentile range of a fit with \texttt{LINMIX} in green.
    We also show the 16th-84th percentile range of a linear fit with constant unit slope in blue 
    Note that the $L_{\rm H\alpha}$ measurements formally have uncertainties, but they are not visible behind the points at this scale.  The annotations, matched to the colors of the errorbars, give the median and 16th-84th percentiles of the fitted model parameters ($y = a_0 + a_1 x$), the intrinsic scatter $\sigma$, and the corresponding Pearson (P) and Spearman (S) statistics and $p-$values. We omit the H II region scale measurements for clarity: the $L^{\rm H~II}_{\rm H\alpha}$ measurements are offset by 0.10 dex lower than the 1.5 kpc scale $L_{\rm H\alpha}$ (i.e. to the left in this plot) with similar correlation, while the $\langle L^{\rm H~II}_{\rm H\alpha}\rangle$ measurements are essentially uncorrelated with $L_X$.
    }
\end{figure}

\begin{figure}
    \centering
    \includegraphics[width=0.50\textwidth]{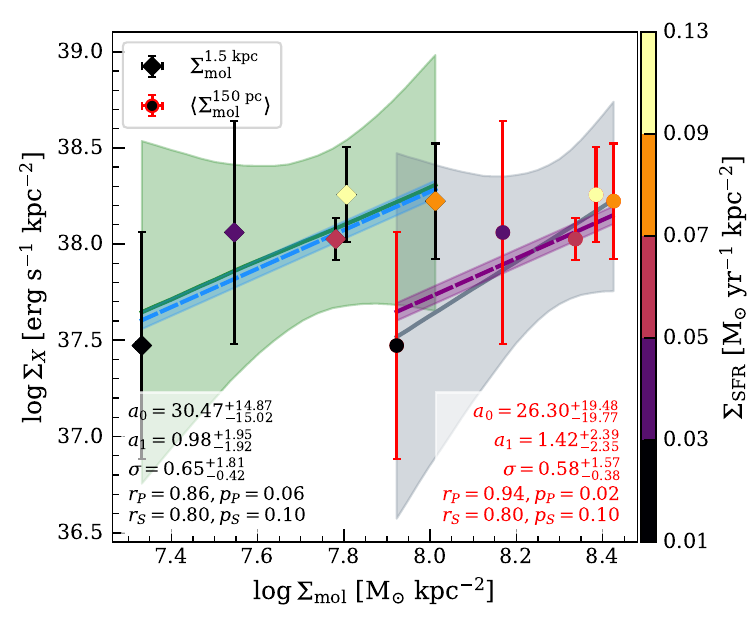}
    \caption{\label{fig:coldgas_scaling} Scaling relations for $\Sigma_X$ with the molecular gas surface density $\Sigma_{\rm mol}$ derived from the ALMA CO(2-1) observations, measured on 1.5 kpc and 150 pc scales and averaged over the \SFRD\ bins in which we extracted X-ray spectra. We show the 1.5 kpc-scale $\rm H\alpha$ surface brightness as diamond markers with red errorbars and the cloud-scale average surface brightness as circles with red errorbars. The face color of the markers corresponds to the \SFRD\ of the bin. We show the pointwise median and 16th-84th percentile range of a \texttt{LINMIX} fit in green for the 1.5 kpc measurements and gray for the 150 pc measurements (note that the uncertainties on $\Sigma_{\rm mol}$ are not visible behind the points). We also show the 16th-84th percentile range of a linear fit with constant unit slope in blue and purple for the 1.5 kpc and 150 pc measurements, respectively; the dashed lines mark the medians.
    The annotation shows the median and 16th-84th percentiles of the fitted model parameters ($y = a_0 + a_1 x$), the scatter $\sigma$, and the corresponding Pearson (P) and Spearman (S) statistics and $p-$values. 
    }
\end{figure}

To compare the ionized and molecular ISM to the properties of the X-ray emitting gas that we measured from spectral fitting, we aggregated the multi-scale H$\alpha$ luminosity and molecular gas surface density into the \SFRD\ bins described in \autoref{sec:spec}. We show the comparison between the 1.5 kpc scale and 150 pc scale measurements and the X-ray surface brightness for H$\alpha$ surface brightness and molecular gas surface density in \autoref{fig:muse_scaling} and \autoref{fig:coldgas_scaling}, respectively. We see that on 1.5 kpc scales, the absorption-corrected luminosity of the X-ray emission is not strongly correlated with the H$\alpha$ surface brightness, significant only at the $90\%$ confidence level by the Spearman metric. This large-scale measurement of $L_{\rm H \alpha}$ must contain contributions from the diffuse ionized gas (DIG) component of the galaxy, which may be less directly connected with the star-forming regions which produce the diffuse X-ray emission. The total H II region luminosities, averaged over each \SFRD\ bin, are systematically slightly lower than the corresponding $L^{1.5~\rm kpc}_{\rm H\alpha}$, with median (maximum) offset 0.10 (0.13) dex, indicating that 80\% (75\%) of the H$\alpha$ luminosity is contributed by the bright H II region candidates in the \citet{groves2023} catalog, with the remaining fraction coming from the DIG and non-detected H II regions. While we saw evidence above for the importance of non-detected star-forming regions in producing the hot ISM, we see here that averaging over the \SFRD\ bins and considering the spectral fitting-derived properties of the X-ray emitting gas washes out this signal. The strength and significance of the correlation between $L^{\rm H~II}_{\rm H\alpha}$ and $L_X$ are not much different from the correlation between $L^{1.5~\rm kpc}_{\rm H\alpha}$ and $L_X$: we compute Spearman statistic $r = 0.80$ and $p-$value $p=0.10$. We see again, however, suggestions that the brightest star-forming complexes alone are less connected to the hot gas X-ray luminosity: for $\langle L^{\rm H~II}_{\rm H\alpha}\rangle$ and $L_X$ we see Spearman statistic $r = -0.30$ and $p-$value $p = 0.62$. We used a Python port\footnote{\url{https://github.com/jmeyers314/linmix}} of the \texttt{LINMIX} algorithm \citep{kelly2007} to fit a linear correlation between $\log L_X$ and measurements of $\log L_{\rm H \alpha}$ to estimate the intrinsic scatter in the data. The scatter in the $\log L^{\rm H~II}_{\rm H\alpha}-\log L_X$ and $\log L^{1.5~\rm kpc}_{\rm H\alpha}-\log L_X$ relations is identical at 0.50 dex, driven by the uncertainty in the absorption-corrected X-ray luminosity due to the large range of $N_H$ compatible with our X-ray spectra. The slopes of these \texttt{LINMIX}-fitted lines are compatible with unity, and so we also fit simple multiplicative scaling relations to $L_X$ as a function of $L_{\rm H \alpha}$, finding
\begin{align}
    \log L_X / L^{1.5~\rm kpc}_{\rm H\alpha} &= -2.28 \pm 0.03,~\sigma=0.18\\
    \log L_X / L^{\rm H~II}_{\rm H\alpha} &= -2.19 \pm 0.03,~\sigma=0.20.
\end{align}
The second relation in particular suggests that we should expect any soft, diffuse emission associated with H II regions to be a factor of $\approx 200$ less luminous than their H$\alpha$ luminosities, consistent with studies of evolved (age $> 5$ Myr) star-forming regions in NGC 2403 by \citet[][see their Table 10]{yukita2010}.

In \autoref{fig:coldgas_scaling}, we see again that aggregating the ISM properties into the \SFRD\ bins in order to compare to the measured properties of the plasma weakens the correlations that we observe when considering the model-independent net counts. The large-scale molecular gas density $\Sigma^{1.5~\rm kpc}_{\rm mol}$ plateaus with \SFRD, creating a pileup in the $\Sigma_X - \Sigma^{1.5~\rm kpc}_{\rm mol}$ plane, though the Spearman statistic still suggests a correlation with $96\%$ confidence. The smaller-scale measurements are not strongly correlated with the X-ray surface brightness. The intrinsic scatter around a line fitted with \texttt{LINMIX} is again essentially the same regardless of the measurement scale, indicating that it is driven by the uncertainty in the X-ray surface brightness. 
\citet{zhang2025} found a $0.67\pm0.23$ slope when fitting the X-ray surface brightness to molecular gas scaling relation in the disk of M51 using $\approx 1.3$ kpc regions, somewhat shallower than the median slope of our \texttt{LINMIX} fit to the 1.5 kpc-scale measurements. However, they benefit from improved constraints on the surface brightness due to the large projected size of M51 and the greater depth of the X-ray data. We again see no evidence in our data of the steep relations they derived for the inner 2 kpc of M51. If we naively fit multiplicative scaling relations between the molecular mass densities, we find 
\begin{align}
    \log \Sigma_X / \Sigma^{1.5~\rm kpc}_{\rm mol} &= 30.35\pm0.03,~\sigma=0.15\\
    \log \Sigma_X / \langle \Sigma^{\rm 150~pc}_{\rm mol}\rangle &= 29.81\pm0.03,~\sigma=0.17.
\end{align}

\begin{figure*}
    \centering
    \includegraphics[width=0.85\textwidth]{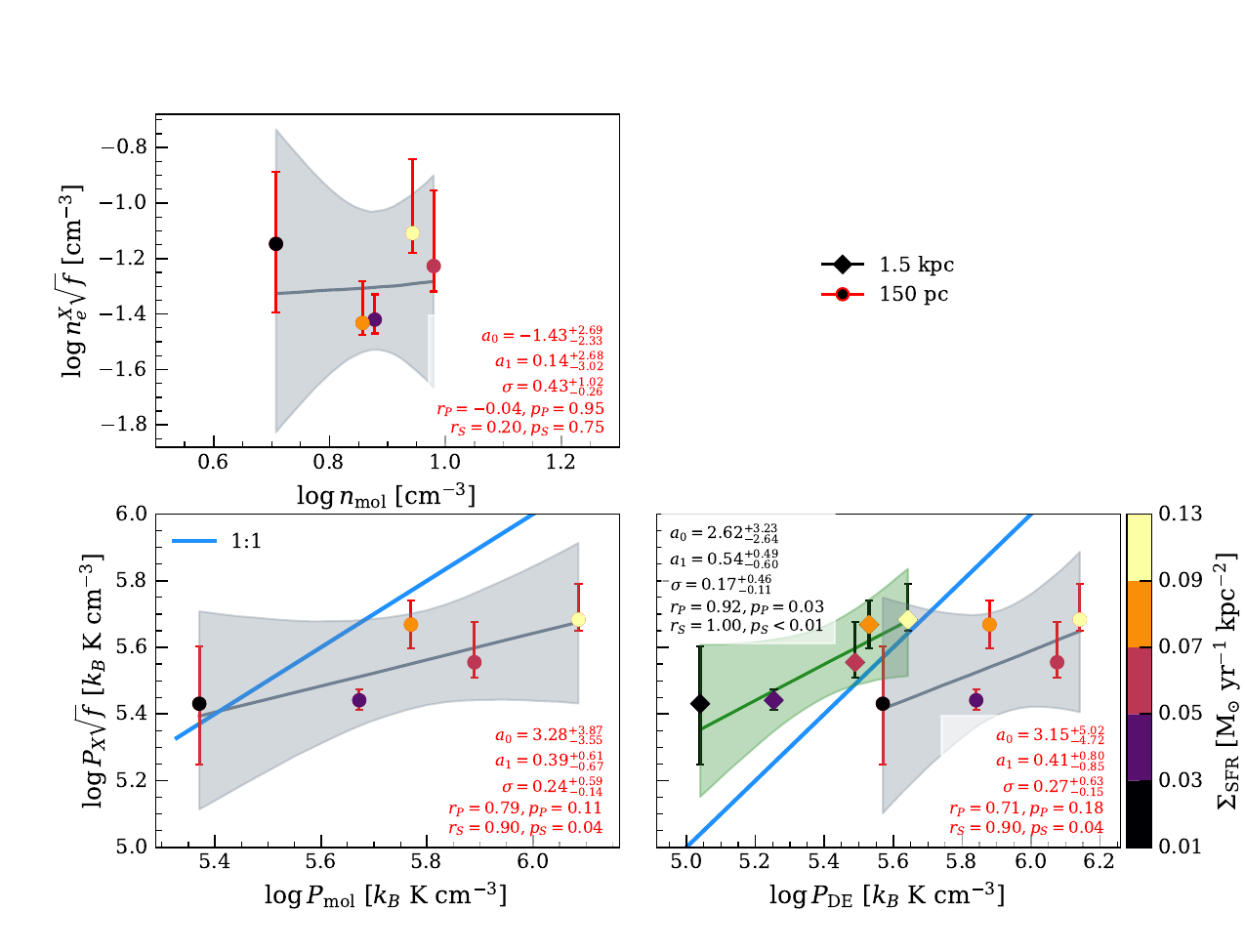}
    \caption{\label{fig:dens_pressure}Number density and pressure comparisons for the hot X-ray emitting and cold molecular phases of the ISM in NGC 4254. The $\rm H_2$ number density was estimated from the molecular mass surface density calculated by \citet{sun2022}, assuming a disk thickness of 200 pc to convert the surface density to volume density. In the lower left panel, we show the turbulent molecular cloud pressure in comparison to the ambient X-ray pressure; in the lower right panel, we show the dynamical equilibrium pressure measured on two different scales in comparison to the X-ray pressure. We show the pointwise median and 16th-84th percentile range of a \texttt{LINMIX} fit in green for the 1.5 kpc measurements and gray for the 150 pc (``cloud-scale'') measurements. 
    Points are colored by their star formation rate density, with brighter, hotter colors corresponding to more intense star formation. In the bottom row, a thick blue line shows the 1:1 correlation between pressures.}
\end{figure*}

In the upper panel of \autoref{fig:dens_pressure} we compare the number density of the averaged molecular cloud complex measured on 150 pc scales to the electron density of the X-ray plasma. We see that the two are uncorrelated: the average density of molecular clouds on small scales is not a predictor of the density of the hot phase of the ISM measured on large scales. In the lower panels of \autoref{fig:dens_pressure}, we compare the pressures in the ISM. We see that the ambient pressure from the hot ISM is typically smaller than the turbulent molecular pressure, with an average offset of 0.23 dex (a factor of 1.7). The offset becomes larger in the highest-\SFRD\ bin, reaching 0.4 dex (a factor of 2.5). The over-pressurization of the turbulent molecular clouds suggests that they are able to expand into the ambient hot medium. However, the offsets are not significantly larger than the uncertainties in the X-ray pressures, and we note again the uncertainty in the volume occupied by the X-ray emitting plasma.

In comparison to the dynamical equilibrium pressure estimated on 1.5 kpc scales (diamond markers in \autoref{fig:dens_pressure}) the hot gas is over-pressurized. \citet{sun2020} found a similar result for the molecular gas, noting that large-scale estimates of $P_{\rm DE}$ may fail to account for the clumpiness of the molecular ISM. They calculated a cloud scale equilibrium pressure (circle markers in \autoref{fig:dens_pressure}), finding that it agrees better with the turbulent cloud pressure. We see that the X-ray emitting plasma is significantly under-pressured when compared to the cloud-scale equilibrium pressure. Under the assumption that the X-ray plasma is volume-filling ($f=1$), the X-ray plasma thus does not appear to contribute strongly to the support of the gas disk on small scales, though it may exceed the dynamical equilibrium pressure on large scales in low \SFRD\ regions of the galaxy. The median offset between the cloud-scale equilibrium pressure and hot ISM pressure is 0.40 dex (a factor of 2.5); if we suppose that the X-ray emitting plasma is instead in dynamical equilibrium with the stellar and gas disk, this implies a hot gas filling factor $f\approx1/6$.

\added{
Our broad takeaways from this multi-scale correlation analysis between ISM phases are as follows. The correlations of H$\alpha$ and molecular gas density with the net X-ray counts are strongest on the largest 1.5 kpc scale we probe, despite the robust constraints that the high-resolution PHANGS datasets give us on the luminosities of H II region complexes and masses of molecular clouds at 150 pc scales. This may indicate that while the most massive, individually detected star forming regions are important in driving the shocks that produce the hot phase of the ISM, smaller non-detected star forming complexes also provide a significant contribution. We are limited in our ability to connect the intrinsic X-ray luminosity of the hot phase of the ISM to the other phases of the ISM on multiple scales due to large X-ray uncertainties, but we see weak evidence for multiplicative scaling relations between the intrinsic X-ray luminosity, H$\alpha$ luminosity, and molecular gas mass. Our limited understanding of the hot gas geometry and filling factor in this face-on galaxy likewise hinder rigorous comparisons of the density and pressure of the phases, but under the assumption of a flat disk-like hot gas geometry we see that the hot gas pressure can exceed the dynamical equilibrium pressure on large scales, indicating that the hot phase of the ISM can expand into the ambient medium.
}

\added{
While our conclusions from this analysis are limited, it is nonetheless apparent that the X-ray emitting hot phase of the ISM is connected to the properties of the other ISM phases as measured on multiple physical scales.
Future studies of the hot ISM in the PHANGS sample, combining X-ray constraints from multiple galaxies, will provide a more definitive picture of these connections between the phases of the ISM on both 1.5 kpc and 150 pc scales.
}

\section{Summary} \label{sec:summary}

We have performed resolved SED fitting to the spiral galaxy NGC 4254 (M99) to derive stellar mass and star formation rate on kpc scales. At the same time, we have produced high-quality point-source subtracted maps of the diffuse X-ray emission, with systematic derivations of the hot gas luminosity and temperature in stacks of 1.5 kpc-scale hexagonal tiles across the galaxy. We find:

\begin{itemize}

    \item There is evidence for a slight increase in the plasma temperature from $\approx 0.20$ keV in the outer, low-\SFRD\ regions of the disk, to $\approx0.27$ keV in more central, higher-\SFRD\ regions, though this trend may be sensitive to our assumption of a prior for the X-ray obscuration based on the H I surface density, estimated from the 21 cm intensity. The temperature measured from a galaxy-integrated spectral fit, $kT = 0.19^{+0.06}_{-0.04}$, is consistent with the lower temperatures measured in the low-\SFRD\ (but physically much larger) outer regions of the galaxy.
    
    \item The hot gas surface brightness increases nearly an order of magnitude from low (0.01$-$0.03) to high (0.09$-$0.13) \SFRD, appearing to plateau in the central region of the galaxy with the most intense star formation. However, the \added{absorption-corrected} hot gas surface brightness (and consequently, the hot gas density) shows somewhat less variation, increasing only be a factor of 2.5 over the same range of \SFRD. The thermodynamic pressure associated with the hot gas, due to its explicit linear dependence on the temperature, increases with the star formation rate surface density \SFRD.
    
\end{itemize}

We combined our X-ray spectral fitting results and resolved SED fitting to interpret the X-ray emission at sub-galactic scales in the context of the simple \citetalias{chevalier1985} model for supernova winds, deriving constraints on the key parameters.

\begin{itemize}
    \item We find that our data are consistent with thermalization efficiencies $\epsilon \approx 50\%$ and mass loading factors $\beta\approx 50-90$, decreasing as $\SFRD^{\approx-1/3}$. \added{Such mass loading factors (corresponding to $\beta_*=\dot M /{\rm SFR} = 1.25-2.25$) are comparable to those seen in the literature on galaxy-integrated scales: \citet{meiksin2016}, for example, found $\dot M /{\rm SFR}$ ranging from $0.5-3$ reproduces the $L_X-{\rm SFR}$ correlation from the \citet{mineo2012} sample.} \added{The (approximate) scaling we recover} is predicted by the upper limit on the mass loading required to keep the plasma from runaway cooling, and is notably the scaling required to make $L_X/{\rm SFR}$ nearly constant with SFR, indicating that the galaxy-integrated $L_X - {\rm SFR}$ correlation may be regulated by mass-loading of the ISM on sub-galactic scales.
    \item Our results are consistent with X-ray production efficiencies of $\eta = 0.5-2\%$. Compared to recent results from M51, we see that our best-fitting model is broadly consistent with the flatter relationship they measured in the outer disk, but our preferred models are unable to produce the negative slopes they observed. Our preferred model produces $\eta \propto \SFRD^{0.34\pm0.17}$, such that the modeled X-ray production efficiency varies by roughly a factor of two across the range of \SFRD\ that we probe.
\end{itemize}

We leveraged the existing high-resolution MUSE and ALMA data from the PHANGS project to investigate multi-scale correlations between the different phases of the ISM.
\begin{itemize}
    \item The net 0.5$-$2 keV counts in any given region of the galaxy are strongly correlated with the H$\alpha$ luminosity and with the total H$\alpha$ luminosity of the H II regions in that region from the \citet{groves2023} PHANGS-MUSE nebular catalog, in line with expectations given the strong correlation between H$\alpha$ luminosity and SFR.
    
    \item However, when we aggregate H$\alpha$ luminosities and molecular gas densities into \SFRD\ bins in order to compare to the plasma properties we derived from spectral fitting, we see that the large-scale H$\alpha$ luminosity and CO-derived molecular gas surface density are not strong predictors of the intrinsic X-ray luminosity and surface brightness, respectively, on the same scale.
    \item We derived simple multiplicative $\Sigma_X-\Sigma_{\rm H\alpha}$ correlations, finding 
    \begin{align}
        \log L_X / L^{1.5~\rm kpc}_{\rm H\alpha} &= -2.28 \pm 0.03,~\sigma=0.18\\
        \log L_X / L^{\rm H~II}_{\rm H\alpha} &= -2.19 \pm 0.03,~\sigma=0.20.
    \end{align}
    for the intrinsic $0.5-2$ keV luminosity $L_X$. The X-ray luminosity in the vicinity of the average star-forming region is a factor of $200\times$ less than its H$\alpha$ luminosity, consistent with results observed for individual star-forming regions in, e.g. NGC 2403 \citep{yukita2010}.
    \item In contrast to recent studies of many resolved regions in M51 by \citet{zhang2025}, we see only a weak relationship between the CO(2-1) emission on 1.5 kpc scales. The power law slope we derive for the CO(2-1)-X-ray scaling relation, $0.98^{+1.95}_{-1.92}$, is not well-constrained, but shows a weak preference for a steeper relation than their result, $0.67 \pm 0.23$.
    \item We see that the thermodynamic pressure of the ambient X-ray plasma is typically smaller than the internal turbulent pressure of molecular clouds. The X-ray pressure can exceed the dynamical equilibrium pressure of the gas and stellar disk of the galaxy measured on large scales, but is significantly smaller than the ``cloud-scale'' equilibrium pressure. The X-ray emitting plasma is thus able to expand into the ambient medium on large scales.
     
\end{itemize}


\begin{acknowledgments}
EBM acknowledges support from Chandra X-ray Center grant GO4-25052B, and from Penn State ACIS Instrument Team Contract SV4-74018 (issued by the Chandra X-ray Center, which is operated by the Smithsonian Astrophysical Observatory for and on behalf of NASA under contract NAS8-03060). This work is based on Chandra ACIS Guaranteed Time Observations (GTO) selected by the ACIS Instrument Principal Investigator, Gordon P. Garmire, currently of the Huntingdon Institute for X-ray Astronomy, LLC, which is under contract to the Smithsonian Astrophysical Observatory via Contract SV2-82024.
B.D.L. and A.A. gratefully acknowledge financial support from the Chandra X-ray Center grant GO4-25052A.
K.B., L.A.L., S.L., and J.A.R. gratefully acknowledge financial support from the Chandra X-ray Center grant AR4-25005X and from the Heising-Simons Foundation grant 2022-3533.
SM is grateful for the grant provided by the National Aeronautics and Space Administration (NASA) through Chandra Award Number GO5-26001X issued by the Chandra X-ray Center, which is operated by the Smithsonian Astrophysical Observatory for and on behalf of NASA under contract NAS8-03060.
S.D. acknowledges support provided by NASA through Hubble Fellowship grant HST-HF2-51551.001-A awarded by the Space Telescope Science Institute, which is operated by the Association of Universities for Research in Astronomy, Inc., for NASA, under the contract NAS 5-26555. 
ARB acknowledges support by NASA under award number 80GSFC24M0006.
JS acknowledges support by the National Aeronautics and Space Administration (NASA) through the NASA Hubble Fellowship grant HST-HF2-51544 awarded by the Space Telescope Science Institute (STScI), which is operated by the Association of Universities for Research in Astronomy, Inc., under contract NAS~5-26555. YHT and ADB acknowledge support from grant NSF-AST 2307441.
KK gratefully acknowledges funding from the Deutsche Forschungsgemeinschaft (DFG, German Research Foundation) in the form of an Emmy Noether Research Group (grant number KR4598/2-1, PI Kreckel) and the European Research Council’s starting grant ERC StG-101077573 (“ISM-METALS"). 
HAP acknowledges support from the National Science and Technology Council of Taiwan under grant 113-2112-M-032-014-MY3.

This work has made use of the ROAR cluster computing facility at Pennsylvania State University.

The collected Chandra data used in this work is available at the Chandra Data Archive, with DOI \dataset[10.25574/cdc.450]{\doi{https://doi.org/10.25574/cdc.450}}.

Based on VLT MUSE observations collected at the European Southern Observatory under 1100.B-0651 (PHANGS-MUSE; PI: Schinnerer). 

This work makes use of ALMA dataset ADS/JAO.ALMA\#2015.1.00956.S.
ALMA is a partnership of ESO (representing its member states), NSF (USA) and NINS (Japan), together with NRC (Canada), MOST and ASIAA (Taiwan), and KASI (Republic of Korea), in cooperation with the Republic of Chile. The Joint ALMA Observatory is operated by ESO, AUI/NRAO and NAOJ.

This work is based in part on observations made with the Karl G. Jansky Very Large Array (VLA; project code: AP206). VLA is operated by the National Radio Astronomy Observatory (NRAO). NRAO is a facility of NSF operated under cooperative agreement by Associated Universities, Inc (AUI). 

This publication uses the data from the AstroSat mission and the UVIT instrument of the Indian Space Research Organisation (ISRO), archived at the Indian Space Science Data Centre (ISSDC). This work is supported by a grant 19ASTROSA2 from the Canadian Space Agency.

We acknowledge the use of public data from the Swift data
archive.

This work is based in part on observations made with the
Galaxy Evolution Explorer (GALEX). GALEX is a NASA
Small Explorer, whose mission was developed in cooperation
with the Centre National d’Etudes Spatiales (CNES) of France
and the Korean Ministry of Science and Technology. GALEX
is operated for NASA by the California Institute of Technology
under NASA contract NAS 5-98034.



This research has made use of the NASA/IPAC Infrared Science Archive (IRSA), which is funded by the 
National Aeronautics and Space Administration and operated by the California Institute of Technology. 
Data used in this work from the 2MASS Large Galaxy Atlas, Spitzer Infrared Nearby Galaxy Survey, 
WISE All-Sky Survey, and KINGFISH are available at IRSA with DOIs 
\dataset[10.26131/IRSA122]{\doi{10.26131/IRSA122}}, 
\dataset[10.26131/IRSA424]{\doi{10.26131/IRSA424}}, 
\dataset[10.26131/IRSA151]{\doi{10.26131/IRSA151 }}, 
and \dataset[10.26131/IRSA71]{\doi{10.26131/IRSA71}}, respectively.

This publication makes use of data products from the Two Micron All Sky Survey, which is a joint project of the University of Massachusetts and the Infrared Processing and Analysis Center/California Institute of Technology, funded by the National Aeronautics and Space Administration and the National Science Foundation.

Based on observations and archival data obtained with the \textit{Spitzer} Space Telescope, which is operated by the Jet Propulsion Laboratory, California Institute of Technology under a contract with NASA.

This publication makes use of data products from the Wide-field Infrared Survey Explorer, which is a joint project of the University of California, Los Angeles, and the Jet Propulsion Laboratory/California Institute of Technology, funded by the National Aeronautics and Space Administration.

This work uses observations made with ESA \textit{Herschel} Space Observatory. \textit{Herschel} is an ESA space observatory with science instruments provided by European-led Principal Investigator consortia and with important participation from NASA.

\end{acknowledgments}

%

\vspace{5mm}
\facilities{CXO(ACIS), ALMA, VLT(MUSE), GALEX, AstroSat(UVIT), Swift(UVOT), SDSS, IRSA, WISE, Spitzer(IRAC, MIPS), Herschel(PACS, SPIRE)}


\software{\texttt{astropy} \citep{astropy2013,astropy2018,astropy2022},  
          \texttt{ACIS Extract} \citep{broos2010, acisextract},
          \texttt{BXA} \citep{buchner2014}
          \texttt{CIAO} \cite{fruscione2006},
          \texttt{Cloudy} \citep{ferland2013}, 
          \texttt{emcee} \citep{foremanmackey2013},
          \texttt{linmix}~(Python) (Josh Meyers et al.),
          \texttt{lightning.py} \citep{lightningpy},
          \texttt{FastHR} \citep{zou2023},
          \texttt{PyNeb} \citep{luridiana2015},
          \texttt{PyAtomDB} (AtomDB Project) ,
          \texttt{Sherpa} \citep{sherpa2001, sherpa2007}}



\appendix
\section{CC85 Model} \label{sec:cc85}

Following \citetalias{chevalier1985} we will take the energy and mass released by a single CCSNe as $E_{SN} = 10^{51}$ erg and $M_{\rm SN} = 3~\rm M_{\odot}$. Writing the number of CCSNe per solar mass formed as $\nu$, we have the energy injection rate into the ISM

\begin{equation} \label{eq:Edot0}
    \dot E = \epsilon \dot E_0 = \epsilon E_{SN} \nu {\rm SFR} = 3.17\times10^{43} \epsilon \nu {\rm SFR}~[{\rm erg~s^{-1}}],
\end{equation}
where the SFR is expressed in $\rm M_{\odot}~yr^{-1}$ and the dimensionless factor $\epsilon \in [0,1]$ accounts for the thermalization efficiency of the SNe shocks.
We can similarly parameterize the mass injection rate into the ISM:

\begin{equation} \label{eq:Mdot}
    \dot M = \beta \dot M_0 = \beta M_{\rm SN} \nu {\rm SFR} = 3 \beta \nu {\rm SFR}~[{\rm M_{\odot}~yr^{-1}}],
\end{equation}
where $\beta \ge 1$ is the mass loading factor, accounting for the entrainment of matter into the ISM by the wind. Note that in some literature, the mass loading factor is defined as the ratio between the mass injection rate and the star formation rate, such that $\beta_* = \dot M / {\rm SFR} = \beta M_{\rm SN} \nu$.

\citet{meiksin2016} derived analytical predictions for the temperature and density of a plasma resulting from the steady-state (i.e., ignoring heat conduction and gravitational cooling) solution of the \citetalias{chevalier1985} model. Most of the emission originates on scales $<R$, the size of the idealized spherical star-forming region, where the temperature and density are roughly constant \citep[see][Appendix A]{meiksin2016}. The temperature is

\begin{equation} \label{eq:kT}
    k T = \frac{2 \bar m \dot E}{5 \dot M} =  \frac{2 \bar m \epsilon \dot E_0}{5 \beta \dot M_0} = 41.02 \frac{\epsilon}{\beta}~[{\rm keV}] ,
\end{equation}
where $\bar m$ is the mean mass per particle ($\bar m = m_p \mu$ for proton mass $m_p$ and reduced mass $\mu$). The model notably predicts no explicit dependence of the temperature on the SFR, such that any SFR-dependence of $kT$ would indicate SFR-dependence of $\epsilon$ and $\beta$. To produce temperatures on the order of $0.1-1$ keV as commonly derived from X-ray spectral fits to nearby galaxies, we must have $-2.6 \lesssim \log \epsilon / \beta \lesssim -1.6$. 

\citet{meiksin2016} derive the central density of the gas as 

\begin{equation}\label{eq:rhoctr}
    \rho_0 = 0.2960 \dot M^{3/2} \dot E^{-1/2} R^{-2},
\end{equation}
corresponding to 

\begin{equation}\label{eq:nhctr}
    n_{H,0} = 0.2960 \frac{X}{m_p} \dot M^{3/2} \dot E^{-1/2} R^{-2},
\end{equation}
where $X$ is the mass fraction of Hydrogen. If we assume that the temperature and density of the steady-state wind are approximately constant inside the wind-launching radius $R$, we can write the luminosity of the gas as

\begin{equation}
    L_{E_1 - E_2} = n_e n_H \Lambda_{12} V = \frac{4}{3} \pi n_e n_H R^3 \Lambda_{12},
\end{equation}
where $\Lambda_{12}$ is the cooling function for the $E_1 - E_2$ bandpass, evaluated at the temperature of the gas. We note the implicit assumption here that the X-ray emitting plasma fills the volume $V$. If we assume $n_e \approx n_H$ and substitute \autoref{eq:nhctr} into the above, we find 

\begin{equation}\label{eq:LXSFR}
    L_{E_1 - E_2} = 9.061\times10^{38}~X^2 \beta^3 \epsilon^{-1} R_{100}^{-1} (\nu {\rm SFR})^2 \frac{\Lambda_{12}}{10^{-23}}~[{\rm erg~s^{-1}}]
\end{equation}
for $R_{100} = R / 100~{\rm pc}$, with $\Lambda_{12}$ in cgs units. If we suppose that $\epsilon$ is constant with SFR and likewise $T$ is not a function of SFR (or that $\Lambda_{12}$ is only weakly a function of $T$ in the range of temperature seen in the hot-phase ISM), then the required scaling to produce $L_X / {\rm SFR} \sim {\rm constant}$ is $\beta \propto {\rm SFR}^{-1/3}$. We note that a reduction in the filling factor of the X-ray plasma would require a larger $\beta$, smaller $\epsilon$, or (counter-intuitively) smaller $R$ to produce the same luminosity at a given SFR.

If the density becomes sufficiently large that the total radiative cooling rate $n_e n_H \Lambda V$ exceeds the energy injection rate $\dot E$, the wind fluid will experience significant radiative cooling, such that the adiabatic assumption is no longer valid. We thus require $\dot E$ to exceed the radiative cooling rate, which produces a corresponding upper limit on the mass loading factor:

\begin{equation}\label{eq:betamax}
    \beta < 32.7 X^{-2/3} \epsilon^{2/3} R_{100}^{1/3} (\nu {\rm SFR})^{-1/3} \left( \frac{\Lambda}{10^{-23}} \right)^{-1/3},
\end{equation}
where the precise dependence of the upper limit on $\epsilon$, $R_{100}$, $\rm SFR$ depends on the assumed $\Lambda$. Using analytic cooling functions, \citet{meiksin2016} derived $\beta_{\rm max} \propto \epsilon^{0.73} ({\rm SFR} / R_{100})^{-0.27}$ and \citet{zhang2014} found a similar result using a cooling function including only free-free emission: $\beta_{\rm max} \propto \epsilon^{0.6} ({\rm SFR} / R_{100})^{-0.4}$. 

\section{Multiwavelength SED Map Fit Quality}\label{sec:quality}
\begin{figure*}
    \centering
    \includegraphics[width=0.95\textwidth]{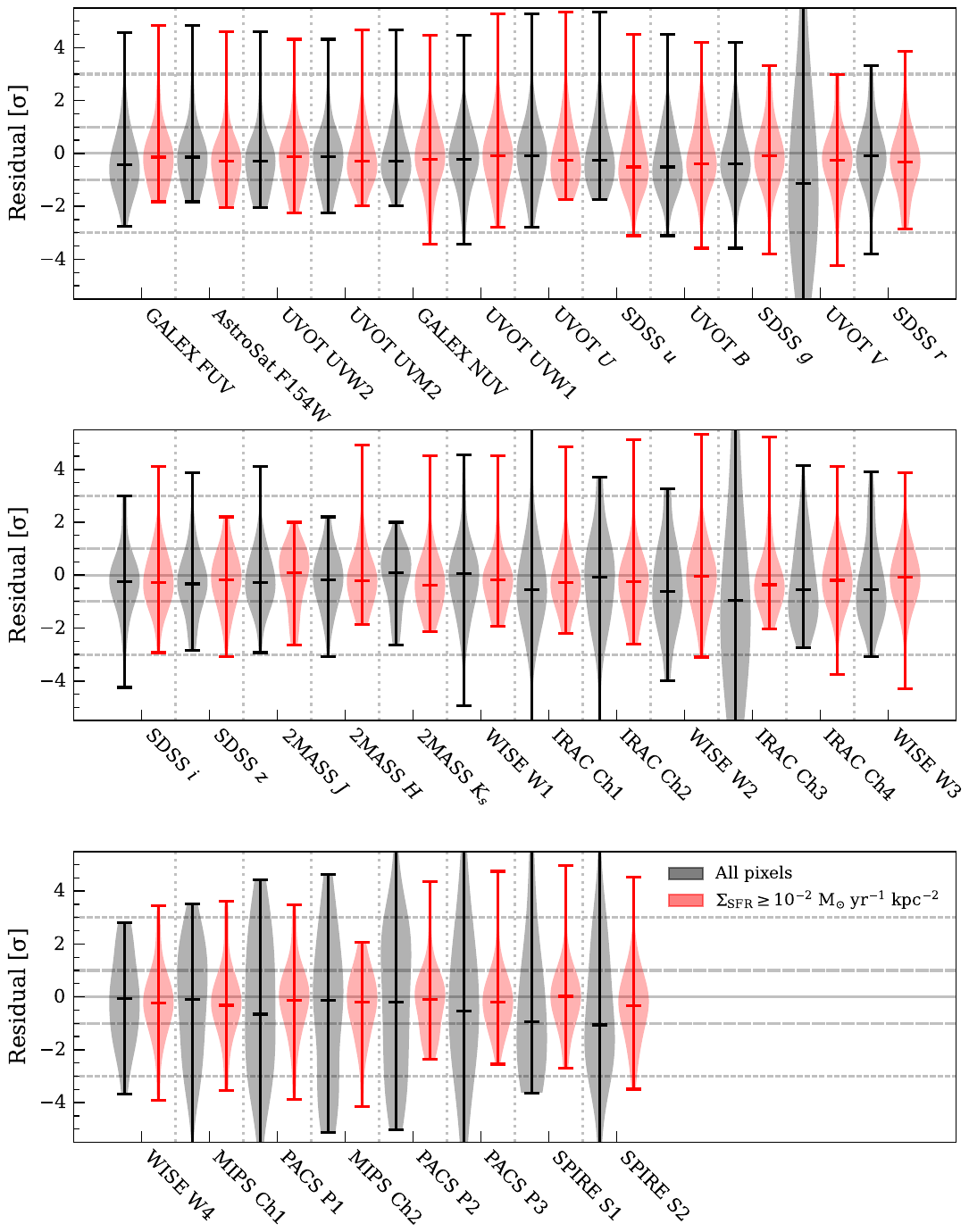}
    \caption{\label{fig:sed_resid} Violin plots of the normalized residual (data$-$model $/ \sigma$) for each bandpass used to fit the multiwavelength SED map. Black violins show all pixels, while red violins show the pixels inside the SFRD $\geq 10^{-2}~{\rm M_{\odot}~yr^{-1}~kpc^{-2}}$ contour. Horizontal lines are plotted at $0,~\pm 1$, and $\pm 3$ to guide the eye.}
\end{figure*}

In \autoref{fig:sed_resid} we show the normalized residuals of all the pixels in our pixel-by-pixel SED fit, and the subset to which we restrict our analysis of the relationship between the X-ray ISM properties and the SFR. We find no strong systematic offsets in the residuals of the dataset, and that the residuals of the pixels inside the SFRD $\geq 10^{-2}~{\rm M_{\odot}~yr^{-1}~kpc^{-2}}$ contour are strongly clustered around 0, with reduced spread (especially in the near- and mid-infrared) due to the larger signal-to-noise in this more intensely star-forming region of the galaxy. We take this to indicate the reliability of our SED fits, and thus our SFR measurements, especially in this region.

\begin{figure*}
    \centering
    \includegraphics[width=0.95\textwidth]{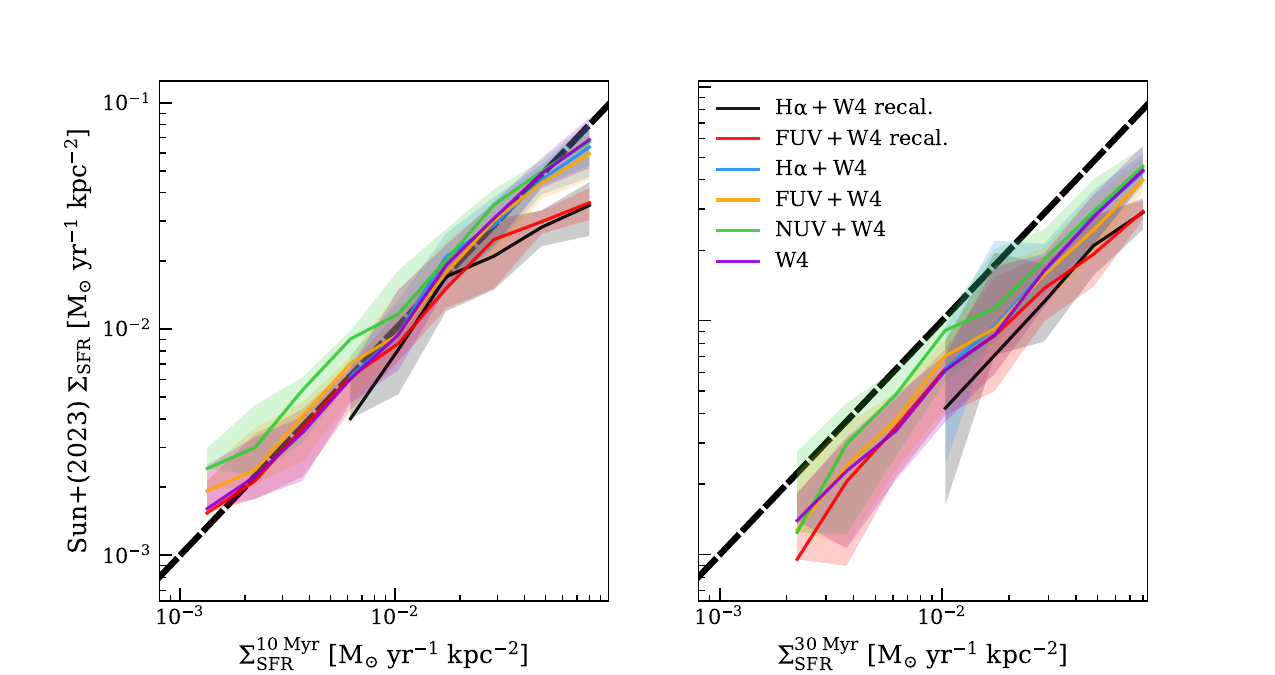}
    \caption{\label{fig:sfrd_comparison} Comparisons between our SED fitting-derived \SFRD\ measurements and the estimators compiled in the high-level \citet{sun2023} catalogs, where W4 = WISE W4 (22 \micron), FUV = GALEX FUV (154 nm), and NUV = GALEX NUV (231 nm). The UV+IR SFR estimators are described by \citet{sun2022} and \citet{leroy2021a}, and the $\rm H\alpha$-based estimators are described by \citet{emsellem2022}. We show the running median and 16th-84th percentile range of the estimators for each hexagonal tile, as a function of the \SFRD\ derived from our maps on two different timescales, 10 and 30 Myr. The thick dashed line shows the 1:1 relation. Agreement is good on the timescales probed by the UV indicators, though as expected our 10 Myr \SFRD\ is systematically larger than the $\rm H\alpha$-derived measurements, which probe star formation on even shorter timescales. Our 30 Myr \SFRD\ measurement, probing a longer timescale than any of the estimators in \citet{sun2023}, is systematically larger by a factor of 2-3. The SFH of most of the regions in the galaxy peaks at stellar ages older than 10 Myr.}
\end{figure*}

We also compare our nonparametric SFH map-derived measurements of \SFRD\ to measurements in the high-level PHANGS measurement tables \citep{sun2023}, which compile a variety of SFR estimators based on combinations of $\rm H\alpha$, UV, and NIR indicators, in  \autoref{fig:sfrd_comparison}. We see good agreement between the short-timescale \SFRD\ estimators in the \citet{sun2023} catalog and our shortest-timescale \SFRD\ measurement from $0-10$ Myr. The \SFRD\ measurement we use throughout the paper, from $0-30$ Myr, is systematically larger than the $\rm H\alpha$, UV, and NIR indicators in part due to our assumption of a flexible SFH shape: the SFH peaks at stellar ages older than 10 Myr, such that the average SFR from 0-30 Myr is systematically larger.

\begin{figure*}
    \centering
    \includegraphics[width=0.95\textwidth]{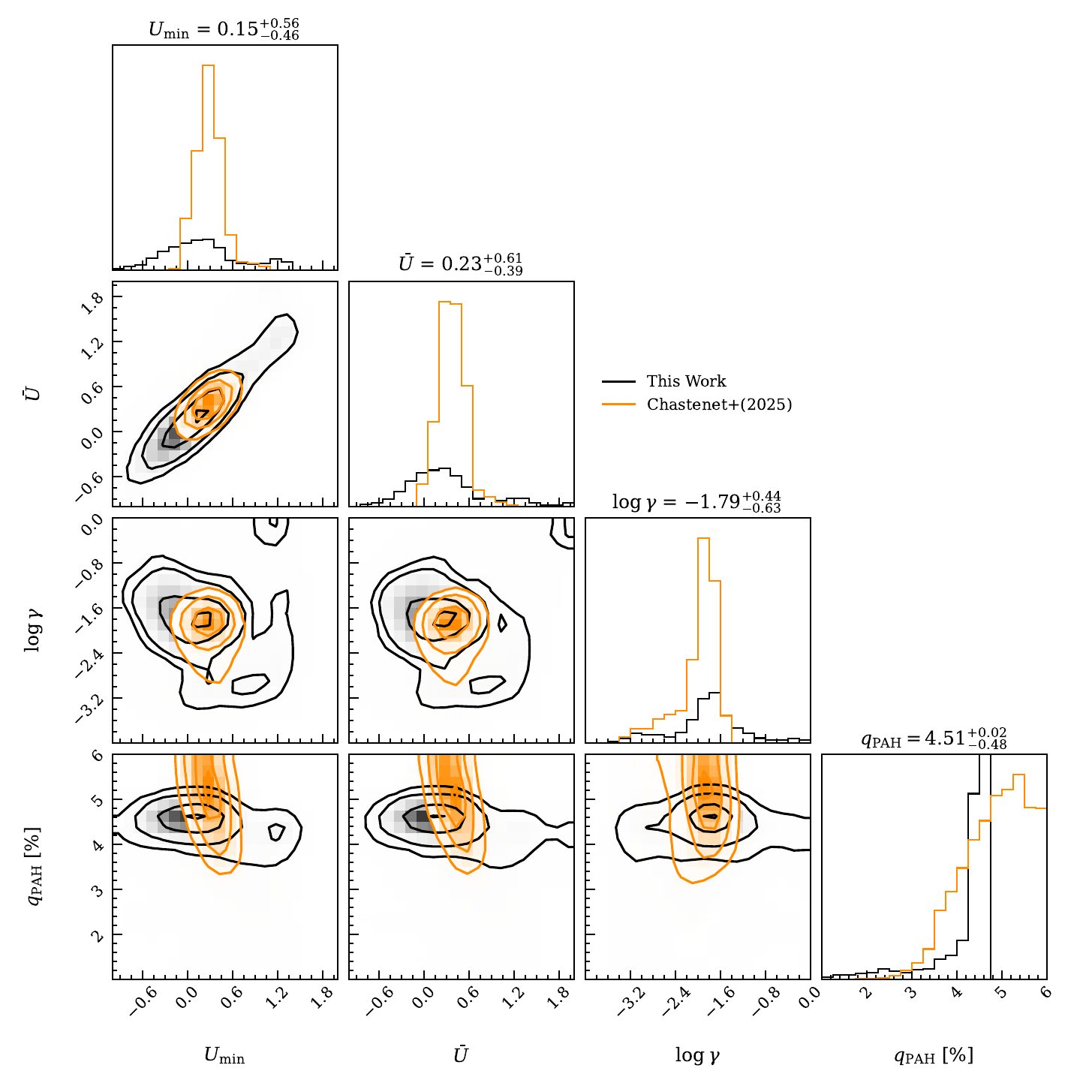}
    \caption{\label{fig:dust_comparison}We show a corner plot of the dust parameters from our SED fits (plus the mean radiation field $\bar U$, derived from $U_{\rm min}$, $U_{\rm max}$, and $\gamma$) in comparison to the values found by \citet{chastenet2025}. We see good agreement between the two sets of dust SED fits, with the notes that \citet{chastenet2025} recover narrower distributions, due to their requirements of a Hershel SPIRE 250 \micron\ signal-to-noise ratio $>1$, and that their implementation of the \citet{draine2007} model allows greater values of $q_{\rm PAH}$, which our fits cap at 4.58\%.}
\end{figure*}

\citet{chastenet2025} performed pixel-by-pixel far-IR SED fits to PHANGS galaxies at comparable resolution to our own UV-to-IR SED fits, using a slightly different implementation of the same \citet{draine2007} dust SED model. Across all 1203 pixels we fit in NGC 4254, we found $\log U_{\rm min} = 0.15^{+0.56}_{-0.46}$, $\log \bar U = 0.23^{+0.61}_{-0.39}$, $\log \gamma = -1.79^{+0.44}_{-0.63}$, and $q_{\rm PAH} = 4.51^{+0.02}_{-0.48}\%$, \added{compatible with the \citet{chastenet2025} estimates of the same parameters across the galaxy:\footnote{The \citet{chastenet2025} parameter maps are available at IRSA with DOI \dataset[10.26131/IRSA581]{https://www.ipac.caltech.edu/doi/irsa/10.26131/IRSA581}} $\log U_{\rm min} = 0.27^{+0.18}_{-0.15}$, $\log \bar U = 0.36^{+0.17}_{-0.17}$, $\log \gamma = -1.90^{+0.19}_{-0.44}$, $q_{\rm PAH} = 5.27^{+1.04}_{-1.03}$.} We note that in their implementation of the \citet{draine2007} models the PAH fraction is allowed to vary over a slightly larger range, whereas our model caps $q_{\rm PAH}$ at 4.58\%, and that we do not fit a dust mass or stellar density due to our assumption of energy balance between the dust and the stellar population attenuation. We compare the distributions of the recovered dust parameters across all pixels in the galaxy in \autoref{fig:dust_comparison}.

\section{X-ray Spectral Fits} \label{sec:specfits}
\begin{figure*}
    \centering
    \includegraphics[width=0.85\textwidth]{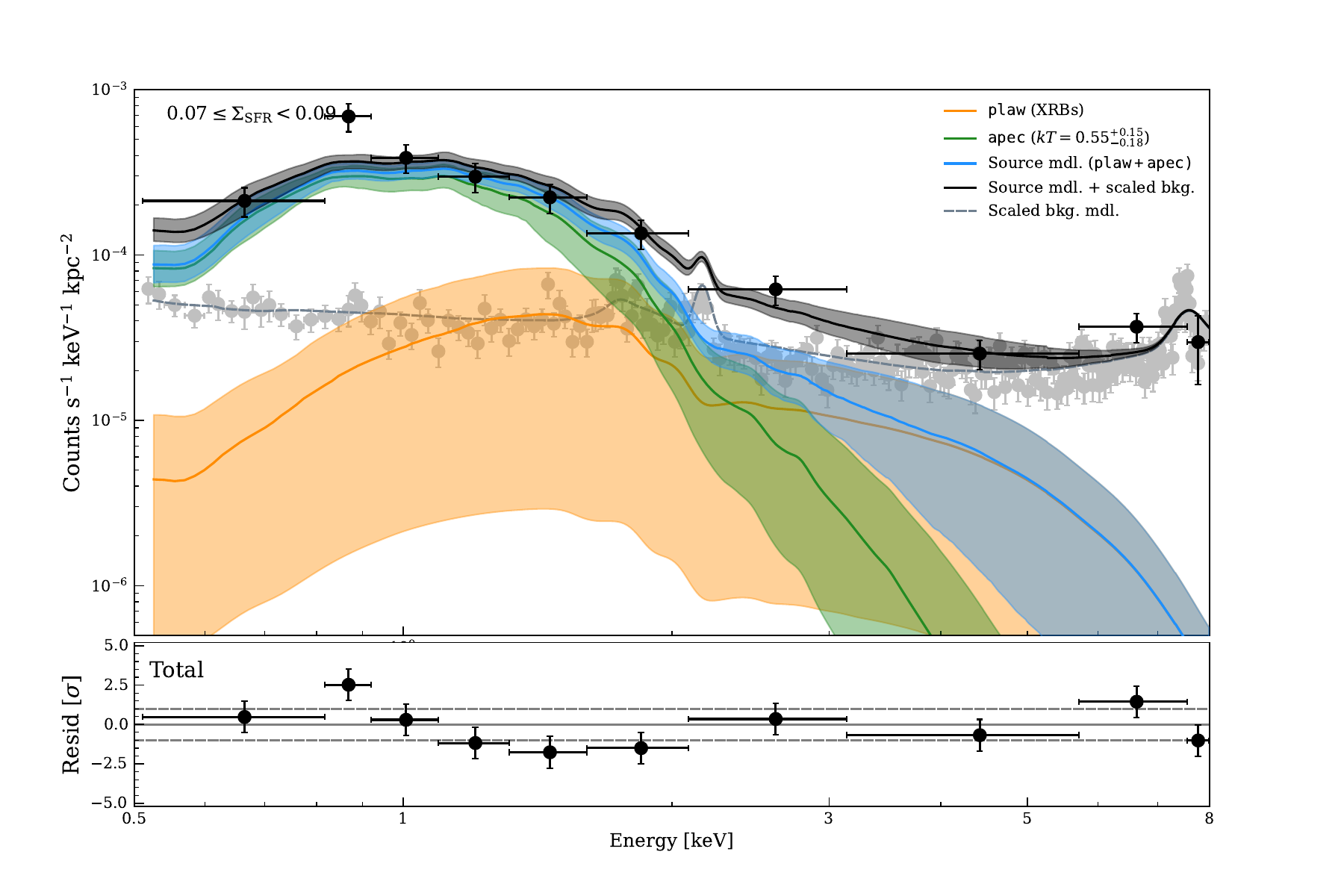}
    \includegraphics[width=0.85\textwidth]{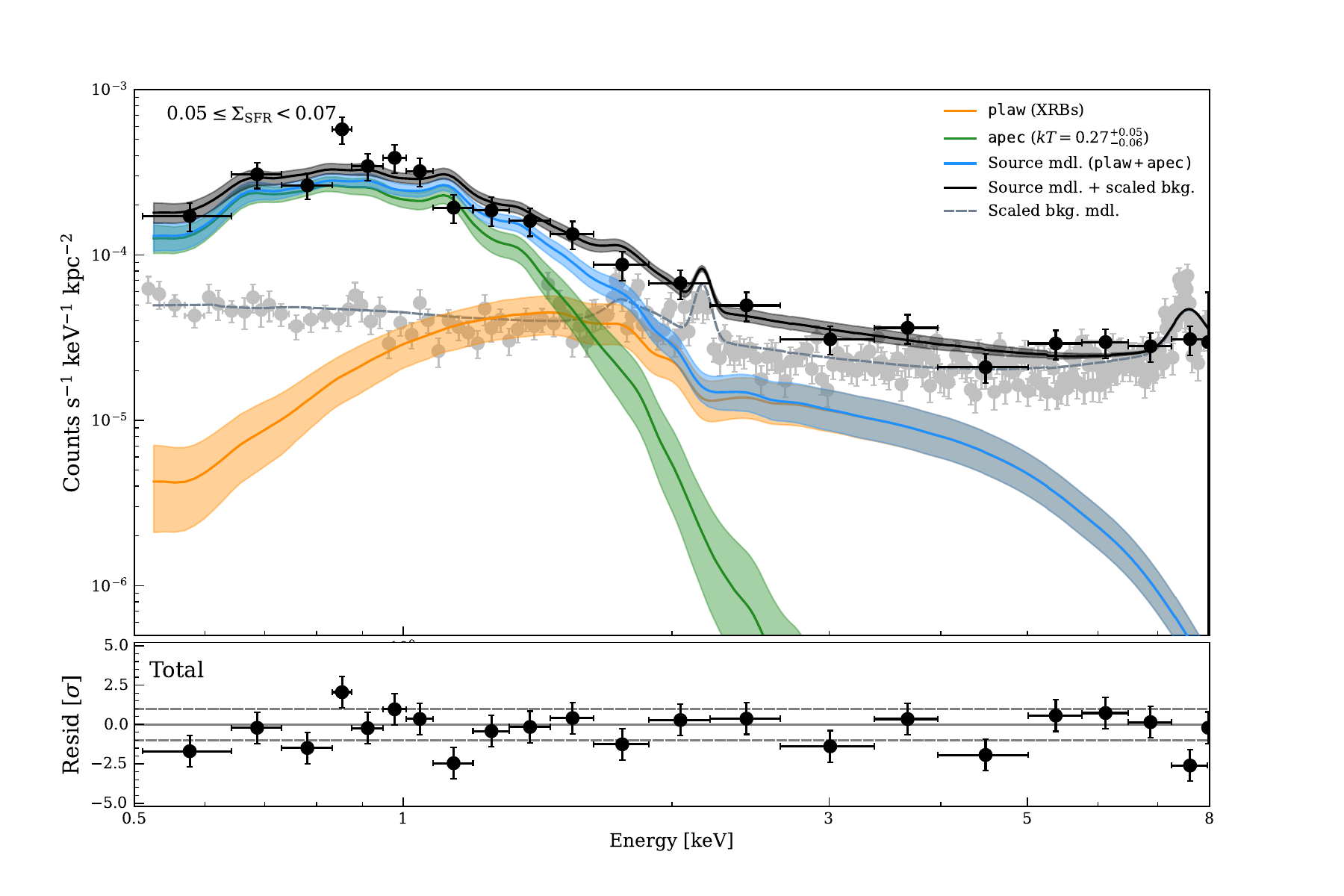}
    \caption{\label{fig:specfits_extra_1}Same as \autoref{fig:xrayfits}, for the $0.07 \leq \SFRD/(\rm M_{\odot}~yr^{-1}~kpc^{-2}) < 0.09$ and $0.05 \leq \SFRD/(\rm M_{\odot}~yr^{-1}~kpc^{-2}) < 0.07$ bins.}
\end{figure*}

\begin{figure*}
    \centering
    \includegraphics[width=0.85\textwidth]{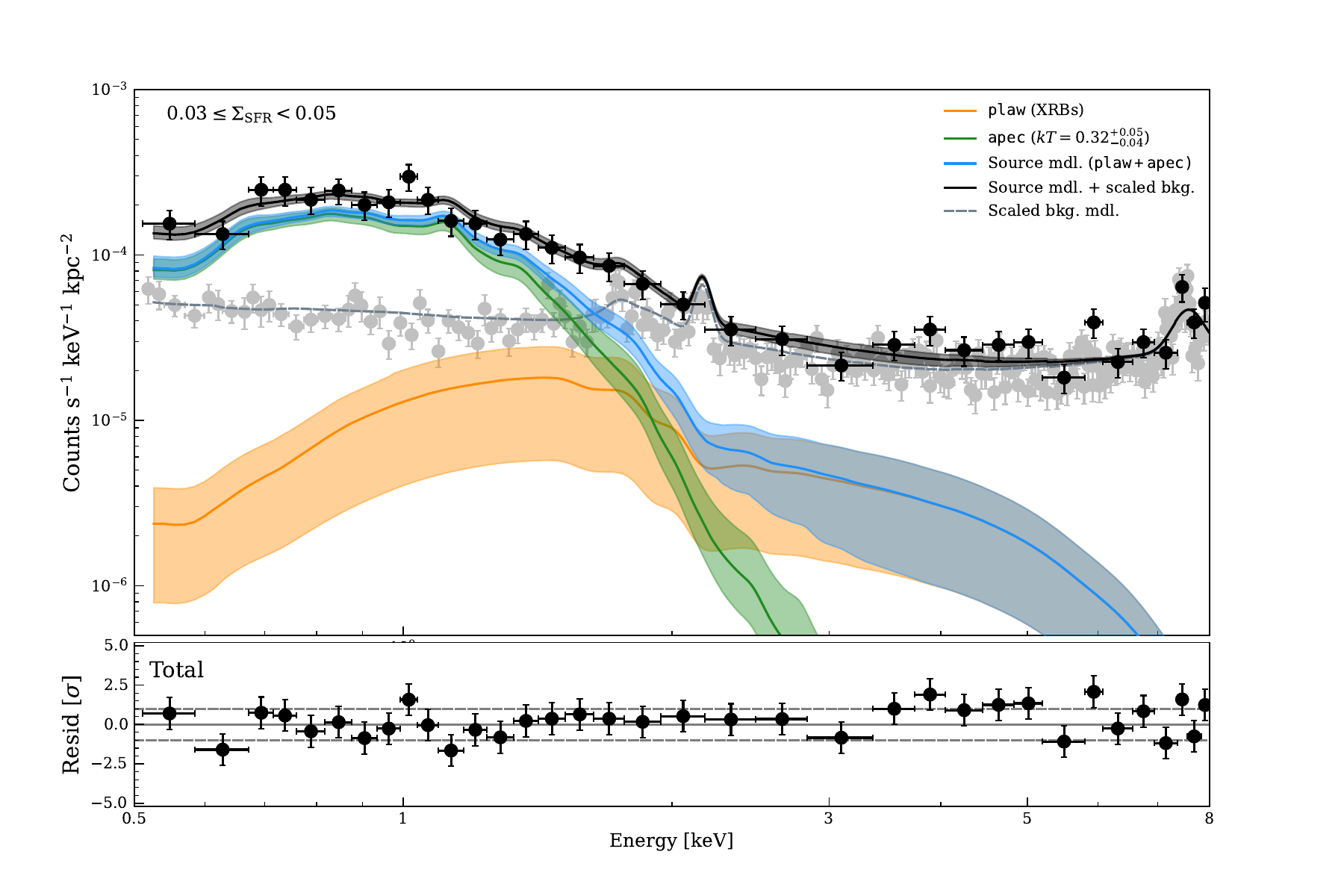}
    \caption{\label{fig:specfits_extra_2}Same as \autoref{fig:xrayfits}, for the $0.03 \leq \SFRD/(\rm M_{\odot}~yr^{-1}~kpc^{-2}) < 0.05$ bin.}
\end{figure*}

In \autoref{fig:specfits_extra_1} and \autoref{fig:specfits_extra_2} we show the X-ray spectral fits to the remaining stacks.

\begin{figure*}
    \centering
    \includegraphics[width=0.85\textwidth]{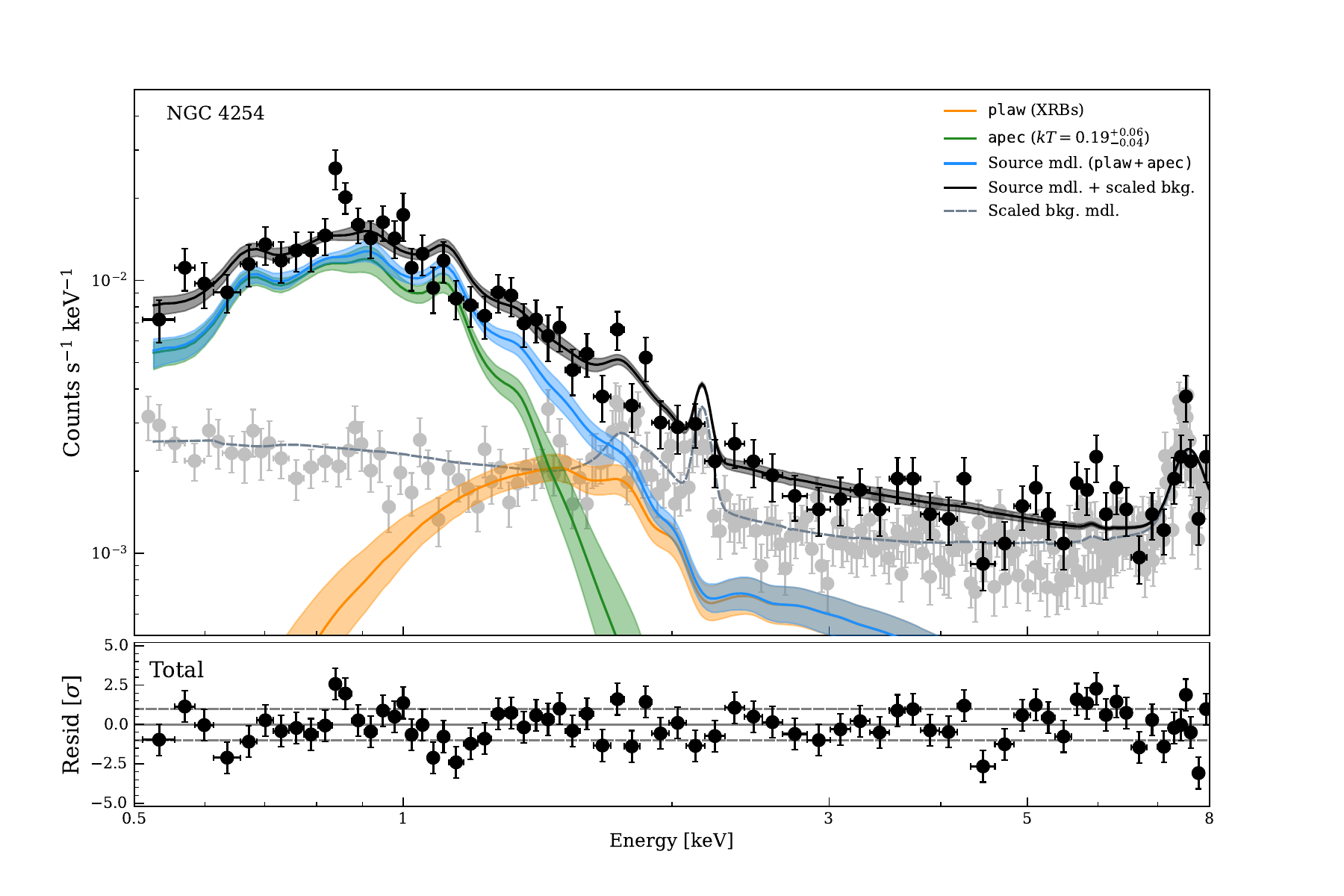}
    \caption{\label{fig:specfits_extra_3}\added{Same as \autoref{fig:xrayfits}, for the integrated spectrum extracted inside the X-ray extent of the galaxy (see \autoref{sec:galint}). Note that for this plot we have chosen not to normalize the $y-$axis by the area over which the spectrum was extracted.}}
\end{figure*}

\section{H I Image} \label{sec:21cm_image}
\begin{figure*}
    \centering
    \includegraphics[width=0.75\textwidth]{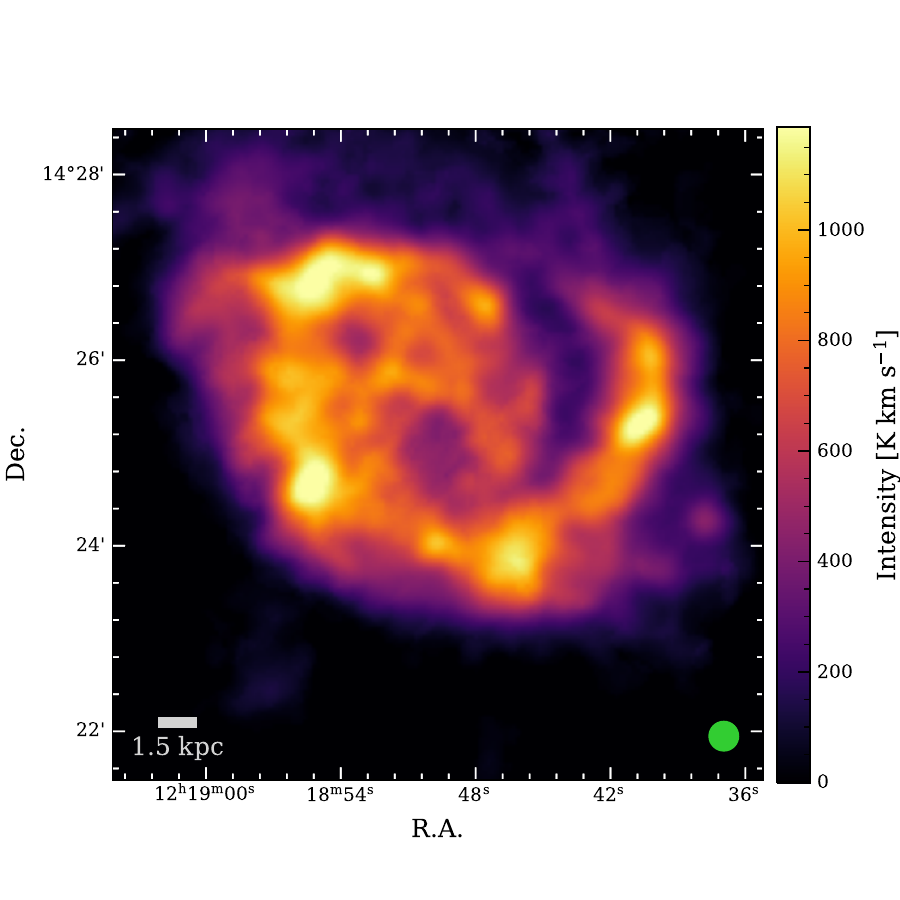}
    \caption{\label{fig:21cm_image} \added{The 21 cm intensity map in K km s$^{-1}$. The beam of the observation is shown in the lower right panel, and the $7'\times7'$ image is centered on the same position as the images displayed in \autoref{fig:ISM_composite} and \autoref{fig:sidebyside}.}}
\end{figure*}

\added{In \autoref{fig:21cm_image} we show the 21 cm map of NGC 4254, which \citet{sun2023} used to produce the atomic gas density measurements we use to derive a prior on the Hydrogen column density for our X-ray spectral fits.}

\bibliography{NGC4254}{}
\bibliographystyle{aasjournalv7}



\end{document}